\RequirePackage{lineno} 

\documentclass[twocolumn,showpacs,preprintnumbers,amsmath,amssymb,nofootinbib]{revtex4}

\usepackage{graphicx}
\usepackage{dcolumn}
\usepackage{bm}

 \usepackage{epstopdf}
\newenvironment{color}[3]{
}{
}

\def\bea{\begin{eqnarray}}
\def\eea{\end{eqnarray}}

\def\pp{\mbox{$p$-$p$}}
\def\auau{\mbox{Au-Au}}

\def\pbpb{\mbox{Pb-Pb}}
\def\aa{\mbox{$A$-$A$}}
\def\nn{\mbox{$N$-$N$}}

\def\pt{$p_t$}

\begin{document} 

\setpagewiselinenumbers
\modulolinenumbers[5]

\preprint{Version 1.8}


\title{Challenging claims of ``elliptic flow'' by comparing a nonjet azimuth quadrupole with jet-related angular correlations from Au-Au collisions at $\sqrt{s_{NN}} = $ 62 and 200 GeV}

\author{Thomas A.\ Trainor}\affiliation{CENPA 354290, University of Washington, Seattle, Washington 98195}
\author{David T.\ Kettler}\affiliation{CENPA 354290, University of Washington, Seattle, Washington 98195}
\author{Duncan J.\ Prindle}\affiliation{CENPA 354290, University of Washington, Seattle, Washington 98195}
\author{R.\ L.\ Ray}\affiliation{Department of Physics, University of Texas at Austin, Austin, Texas 78712}


\date{\today}

\begin{abstract}
%
{\bf Background:} 
A component of azimuth correlations from high-energy heavy ion collisions varying as $\cos(2\phi)$ and denoted by symbol $v_2$ is conventionally interpreted to represent ``elliptic flow,'' a hydrodynamic manifestation of the initial-state \aa\ overlap geometry. Several numerical methods are used to estimate $v_2$, resulting in various combinations of ``flow'' and ``nonflow'' that reveal systematic biases in the $v_2$ estimates. QCD jets contribute strongly to azimuth correlations and specifically to the $\cos(2\phi)$ component. 
{\bf Purpose:} 
We question the extent of jet-related (``nonflow'') bias in and hydrodynamic ``flow'' interpretations of $v_2$ measurements.
{\bf Method:} 
We introduce two-dimensional (2D) model fits to angular correlation data that distinguish accurately between jet-related correlation components and a {\em nonjet azimuth quadrupole} that might represent ``elliptic flow'' if that were relevant. We compare measured jet-related and ``flow''-related data systematics and determine the jet-related contribution to $v_2$ measurements.
{\bf Results:} 
Jet structure does introduce substantial bias to conventional $v_2$ measurements, making interpretation difficult. The nonjet quadrupole exhibits very simple systematics on centrality and collision energy---the two variables factorize. Within a \auau\ centrality interval where jets show no indication of rescattering or medium effects the nonjet quadrupole amplitude rises to 60\% of its maximum value.
{\bf Conclusions:}  
Disagreements between nonjet quadrupole systematics and hydro theory expectations, the large quadrupole amplitudes observed in more-peripheral \auau\ collisions and a significant nonzero value in \nn\ $\approx$ \pp\ collisions strongly suggest that the nonjet quadrupole does not arise from a hydrodynamic flow mechanism. 
\end{abstract}


\maketitle

\section{Introduction}

Measurements of a single Fourier coefficient of the ``azimuthal anisotropy'' of particle momenta in RHIC heavy ion collisions have been interpreted to indicate production of a thermalized QCD medium with low viscosity, frequently invoked as evidence for a ``perfect liquid''~\cite{gyulassy,review}. That conclusion is based on a conventional interpretation of the $v_2 = \langle \cos(2\, \phi)\rangle$ anisotropy component as a measure of elliptic flow, a conjectured hydrodynamic (hydro) response to density and pressure gradients in the initial collision system corresponding to the transverse eccentricity of the \aa\  overlap region~\cite{ollitrault}. In  a hydro context large elliptic flow values combined with other measurements are interpreted to imply large energy densities, rapid thermalization and small viscosities~\cite{hydro,hydro2}.

However, questions persist concerning $v_2$ measurements, their accuracy and their interpretation. Most conventional $v_2$ measurement methods~\cite{poskvol}, denoted here by the term {\em nongraphical numerical methods} (NGNM), do not distinguish accurately between an isolated {\em azimuth quadrupole} ($m  =2$ cylindrical multipole) Fourier component conjectured to represent elliptic flow and ``nonflow''---a catch-all term representing several possible contributions to $v_2$, but mainly the $m=2$ Fourier component of a two-dimensional (2D) peak attributed to jets~\cite{axialci,ptscale,ptedep,lepmini,anomalous}. Whatever the precision of $v_2$ measurements the physical phenomena actually represented by any $\cos(2\phi)$ asymmetry measurement can be questioned~\cite{gluequad,nohydro}.

Conventional quadrupole measures $v_2\{\text{method}\}$ motivated from a hydro context~\cite{ollitrault} are difficult to interpret, and the statistical properties of some $v_2$ methods lead to substantial systematic bias. In Ref.~\cite{flowmeth}  it was shown that event-plane $v_2\{\text{EP}\}$~\cite{poskvol} is a close approximation to two-particle cumulant $v_2\{2\}$~\cite{2004}, in turn equivalent to the $m = 2$ Fourier coefficient of a projection of all 2D angular correlations onto 1D azimuth [see Eq.~(\ref{eq1})], which may include a large contribution from a prominent 2D peak interpreted in a perturbative QCD (pQCD) context as representing minimum-bias jet structure~\cite{porter2,porter3,ppprd,fragevo,anomalous}.  Four-particle cumulant $v_2\{4\}$~\cite{borg,starv24} reduces, but does not necessarily eliminate, the jet-related contribution.\footnote{The mean jet fragment multiplicity increases from $\approx 2.5$ in \pp\ to $\approx 8$ in central \auau\ collisions~\cite{ppprd,hardspec,fragevo}.} Other model-dependent strategies have been invoked in attempts to reduce the ``nonflow'' (jet) contribution to $v_2$, but their effectiveness remains uncertain~\cite{gluequad,multipoles}.

An alternative  method introduced in Ref.~\cite{davidhq} employs physical-model-independent analysis to isolate geometrically a {\em nonjet} azimuth quadrupole from other contributions. Nonjet (NJ) quadrupole amplitudes are obtained from fits to 2D angular correlations on azimuth $\phi$ and pseudorapidity $\eta$. Measurements  of the NJ quadrupole over all \auau\  centralities and a large energy interval provide qualitatively new insights into the quadrupole phenomenon conventionally attributed to elliptic flow.  NJ quadrupole amplitudes obtained with 2D model fits follow simple trends on centrality and energy described by just two initial-state parameters for all systems down to $\sqrt{s_{NN}} \approx 13$ GeV.

The quadrupole analysis method introduced in Ref.~\cite{davidhq} is based on algebraic study of $v_2$ methods in Refs.~\cite{flowmeth,gluequad} and initial experience with 130 GeV data in Refs.~\cite{axialci,hadrogeom} where general model-fit analysis of 2D angular autocorrelations was first introduced. The same model-fit method was refined and elaborated in Ref.~\cite{anomalous} where the primary focus was the energy and centrality systematics of angular correlations attributed to minimum-bias jets or {\em minijets}. The present study combines the numerical results of Refs.~\cite{davidhq,anomalous} to examine the systematic relation between the NJ quadrupole and minijets and to test the validity of the conventional elliptic flow interpretation for the former. 

In this study we examine the distinction between  nonjet and jet-related quadrupole contributions in relation to other correlation structure. We review the centrality and energy dependence of the NJ quadrupole in terms of Glauber-model parameters as reported in Ref.~\cite{davidhq} and contrast those trends with minijet systematics as established in Ref.~\cite{anomalous}. We compare NJ quadrupole results with previous $v_2\{\text{method}\}$ measurements and with hydro expectations. 
We conclude that  NJ quadrupole variations on energy above 13 GeV and all \auau\  centralities are remarkably simple. Those trends and comparisons with minijet systematics appear to contradict conventional hydro expectations for elliptic flow. For example, the NJ quadrupole increases to 60\% of its maximum value within a \auau\  centrality interval where the lowest-energy-jet-related correlations are
consistent with a transparent collision system, as explained below.

This article is organized as follows:
Section~\ref{anmeth}  reviews analysis methods applied to 2D angular correlations and ``flow'' analysis.
Section~\ref{flowjet} summarizes the conflict between jet and flow interpretations of correlation structure that form a larger context for the present study.
Section~\ref{angcorr} presents measured angular correlations and 2D model fits.
Section~\ref{2dfit} reviews model-fit results for jet-related and nonjet quadrupole correlation components and a universal parametrization of energy and centrality dependence for the latter.
Section~\ref{quadjets} compares nonjet quadrupole and jet-related trends in the context of hydrodynamic expectations for the former.
Section~\ref{disc} presents a discussion of selected results, and
Section~\ref{summ} summarizes

\section{Analysis Methods} \label{anmeth}

A major emphasis of this study is accurate distinction between jet-related and nonjet quadrupole components of angular correlations and the energy and centrality systematics of the latter---what those imply for physical interpretation of the nonjet quadrupole phenomenon. We examine the underlying assumptions and systematic uncertainties of the model-fit analysis method in comparison with alternative $v_2$ analysis methods claimed to support flow interpretations.

\subsection{Correlation spaces}

Two-particle correlations are structures in the pair density on 6D momentum space $(p_{t1},\eta_1,\phi_1,p_{t2},\eta_2,\phi_2)$ that deviate from some defined reference density. In this analysis we study $p_t$-integral correlations on angular subspace $(\eta_1,\phi_1,\eta_2,\phi_2)$, where the angle parameters for relativistic collisions are pseudorapidity $\eta$ (related to polar angle $\theta$) and azimuth $\phi$. We can reduce $(\eta_1,\phi_1,\eta_2,\phi_2)$ to a viewable 2D space with no significant loss of correlation information by using {\em angular autocorrelations}~\cite{inverse}. In place of transverse momentum \pt\ one can define transverse rapidity $y_t = \log[(m_t + p_t)/m_h]$, where $m_h$ is a hadron mass, to provide improved  access to low-\pt\ structure.

An {autocorrelation} as conventionally defined is derived from a pair density $\rho(x_1,x_2)$ by averaging along diagonals in space $(x_1,x_2)$ parallel to sum axis $x_\Sigma = x_1 + x_2$. The averaged pair density $\rho(x_\Delta)$ on difference axis $x_\Delta = x_1 - x_2$ is then an autocorrelation~\cite{inverse}. For correlation structure approximately uniform on $x_\Sigma$ (``stationarity''), typical over $2\pi$ azimuth and within a limited pseudorapidity acceptance $\Delta \eta$ centered at the origin, angular correlations remain undistorted~\cite{hadrogeom}.   Within the STAR  time projection chamber (TPC) acceptance~\cite{starnim} {2D angular autocorrelations} are lossless projections of $p_t$-integral two-particle momentum space onto subspace $(\eta_\Delta,\phi_\Delta)$~\cite{flowmeth}.  The $\phi_\Delta$ axis is divided into {\em same-side} (SS, $|\phi_\Delta| < \pi/2$) and {\em away-side}  (AS, $\pi/2 < |\phi_\Delta| < 3\pi/2$) intervals.

\subsection{Correlation measures}

There are several alternatives for the definition of a correlation measure. The basic element is a histogram of covariances representing correlations of event-wise fluctuations between pairs of 2D bins on $(\eta,\phi)$. $\Delta \rho = \rho - \rho_{ref}$ represents a covariance density, where object pair density $\rho$ contains the structure of interest and reference density $\rho_{ref}$ may be defined in terms of a factorization assumption or constructed from mixed-event pairs. 

{\em Per-pair density ratio} $\Delta \rho / \rho_{\text{ref}} = \rho_\text{sib} / \rho_\text{mix} - 1$ (sometimes referred to as a ``correlation function'' and denoted by $C$) varies with system size as $1 / n_{ch}$ ($n_{ch}$ is charge multiplicity) absent other physical changes. $\rho_\text{sib}$ and $\rho_\text{mix}$ represent sibling (same-event) pairs and mixed-event pairs. In previous analysis we introduced a statistical measure whose variation with $n_{ch}$ reflects only nontrivial physical changes in correlations, the {\em per-particle density ratio} $\Delta \rho / \sqrt{\rho_\text{ref}}$ (Pearson's normalized covariance~\cite{pearson1,pearson2} converted to  a density ratio) that exhibits the desired properties, since $\sqrt{\rho_\text{ref}} \propto n_{ch}$~\cite{inverse,ptscale,ptedep,anomalous}. We introduce previous correlation measurements in terms of that measure. We also reconsider what ``particle'' type best serves as a scaling reference in a given context---final-state hadrons as in $\Delta \rho / \sqrt{\rho_\text{ref}}$, initial-state participant nucleons or the number of \nn\ binary collisions---and rescale some of  the correlation data accordingly.

\subsection{Two-dimensional correlation model}

We require a 2D model function that describes all minimum-bias correlation data (no imposed \pt\ cuts) for all collision systems from \pp\ to central \auau\ for RHIC higher energies (i.e.\ 62 and 200 GeV).
Inspection of 2D data histograms reveals that $p_t$-integral pair-density difference $\Delta \rho(\eta_\Delta,\phi_\Delta)$ contains two types of structure: $\eta_\Delta$-dependent 1D and 2D peaks and $\eta_\Delta$-independent sinusoids $\cos(\phi_\Delta)$ and $\cos(2\phi_\Delta)$, where the $\cos(2\phi_\Delta)$ sinusoid (quadrupole) can be related to $v_2$ measurements.
%
%
We therefore define a model of  2D angular correlations that includes a part varying with $\eta_\Delta$ (2D) and a part independent of $\eta_\Delta$ (1D) composed of the $m = 1,\,2$ terms of a (truncated) Fourier series
\bea \label{eq1}
\frac{\Delta \rho}{\sqrt{\rho_{\text{ref}}}} \hspace{-.02in} &\equiv& \hspace{-.02in} \frac{\Delta \rho_\text{2D}}{\sqrt{\rho_\text{ref}}}(\eta_\Delta,\phi_\Delta) \hspace{-.02in} + \hspace{-.02in} 2 \sum_{m=1}^2 \frac{\Delta \rho [m]}{\sqrt{\rho_\text{ref}}} \cos(m \phi_\Delta).
\eea
No higher terms in the Fourier series are {\em required} by the data~\cite{multipoles,sextupole,anomalous}.
 Fourier coefficients $\Delta \rho [m]/\sqrt{\rho_\text{ref}} = V_m^2/2 \pi  n$ include {\em power-spectrum} elements $V_m^2 = \sum_{i \neq j}^{n,n-1} \cos(m[\phi_i - \phi_j]) \equiv \overline{n(n-1) \langle \cos(m \phi_\Delta) \rangle}$~\cite{flowmeth}, where $n$ is the multiplicity in one unit of $\eta$ and $2\pi$ azimuth, so $n/2\pi \approx d^2n_{ch}/d\eta\, d\phi \equiv \rho_0$. Angle brackets denote event-wise means, overlines denote event-ensemble means. The first term $\Delta \rho_{\rm 2D}/\sqrt{\rho_{ref}}$ is a combination of 1D and 2D peaked functions (Gaussians)~\cite{anomalous}. $V_2^2/2\pi n  \equiv  \rho_ 0 v_2\{2D\}^2 \equiv A_Q\{2D\}$ defines quadrupole measure $v_2\{2D\}$ as a variant of conventional $v_2\{\text{method}\}$ measures. The detailed fit model is defined in Sec.~\ref{fitmodel}.
 
 The 2D fit model described above and in Sec.~\ref{fitmodel}  is {\em not based on physical assumptions}, only on the simple structures observed in minimum-bias (no \pt\ cuts) 2D angular correlations common to \pp\ collisions and \aa\ collisions for all centralities at RHIC energies. Subsequent study resulted in identifying two elements of the model with a MB dijet contribution: the SS 2D Gaussian and the AS 1D dipole. The SS 2D Gaussian projected onto 1D azimuth can be represented by a Fourier series including a quadrupole component which is then described as the jet-related quadrupole. The $m=2$ term of Eq.~(\ref{eq1}) is then the {\em nonjet} (NJ) quadrupole. Given that interpretation $v_2\{2D\}$ would coincide approximately with $v_2\{2\}$ in the absence of jets but  continues to measure a nonjet quadrupole (not associated with the SS 2D peak) to the statistical limits of data in the presence of dijets~\cite{gluequad}.  
 
 It is important to note that the truncated Fourier series in Eq.~(\ref{eq1}) is defined over the {\em full} $2\pi$ azimuth interval such that the series terms are orthogonal. Restricting a cosine function to a limited interval (such as away-side $[\pi/2,3\pi/2]$ only) would result in an isolated peak that has its own complex Fourier series representation, substantially complicating the model and producing misleading results. The linear independence of terms in Eq.~(\ref{eq1}) is then compromised. If the first term of the model representing several peaked distributions is omitted from the model the equivalent structure in the data would then contribute to the 1D Fourier series as ``higher harmonics''~\cite{multipoles}, but the curvature on $\eta$ of such terms would be large, in contrast to the unique AS dipole and NJ quadrupole terms with their negligible curvatures (e.g.\ within the STAR TPC acceptance).

\subsection{``Flow'' and ``nonflow''}

Conjectured elliptic flow, a possible hydrodynamic response to initial pressure/density gradients and overlap geometry in non-central \aa\ collisions, is assumed to be measured by the quadrupole ($m=2$) term in a Fourier-series decomposition of the entire final-state azimuth distribution~\cite{poskvol}. The $m=2$ Fourier coefficient is commonly represented by symbol $v_2$. Fourier analysis is applied to all azimuth structure (``anisotropy'') assuming that elliptic flow dominates that structure. However, possible ``nonflow'' (unspecified non-hydrodynamic) contributions to $v_2$ are admitted. A variety of schemes has been introduced to detect and reduce ``nonflow'' bias, but considerable uncertainty remains for conventional $v_2$ methods~\cite{2004}.
Distinctions between ``flow'' and ``nonflow'' have been extensively discussed (e.g., Refs.~\cite{2004,gluequad,multipoles}).

The assumptions that support such terminology can be questioned. Minimum-bias jets contribute strongly to ``azimuthal anisotropy,'' are predicted for high-energy nuclear collisions and must form a substantial contribution to ``nonflow'' especially at lower \pt. And the nonjet contribution to $v_2$ may not in fact be a flow phenomenon. In the present context we refer instead to a nonjet quadrupole (what might represent elliptic flow) and a jet-related quadrupole ($v_2$ contribution mainly from jets and mainly from a prominent SS 2D peak in 2D angular correlations). 

It has been demonstrated that 2D model fits to angular correlations distinguish jet-related structure from the NJ quadrupole with few-percent accuracy~\cite{flowmeth,gluequad,davidhq,davidhq2,anomalous}. The model functions used in the 2D model fits are motivated by empirical observations of data structure, not physical interpretations of structure components~\cite{axialci,ptscale,flowmeth,gluequad,anomalous}.  The accurate separation of NJ quadrupole and jet-related components  by means of 2D model fits and estimates of jet-related bias (``nonflow'') in published $v_2$ measurements obtained with conventional ``flow'' methods are discussed further in the Appendix.

\subsection{Centrality and eccentricity measures} \label{centecc}

Several centrality measures can be related to observed charge multiplicity $n_{ch}$ integrated within some angular acceptance $(\Delta \eta,\Delta \phi)$ based on the Glauber model of \aa\ collisions~\cite{centmeth}. The common element is the fractional cross section $\sigma/\sigma_0 \equiv b^2/b_0^2$ inferred experimentally from the measured minimum-bias event distribution on $n_{ch}$.

Glauber Monte Carlo parameters $N_{part}$ (number of participant projectile nucleons N) and $N_{bin}$ (\nn\ binary collisions) are related to $n_{ch}$ within the STAR TPC acceptance. Centrality measure $\nu \equiv 2\, N_{bin} / N_{part}$ estimates the mean number of N-N {binary collisions} per participant pair (mean participant path length). We retain the same 200 GeV Glauber parameters for all energies as purely geometrical measures of \aa\ centrality (with 200 GeV N-N cross section $\sigma_{NN} = 42$ mb assumed). 

It is conventionally assumed that elliptic flow represented by $v_2$ is a manifestation of the azimuth asymmetry of the initial \aa\ overlap geometry, specifically the eccentricity denoted by $\epsilon$. The \aa\ eccentricity is estimated by Glauber-model simulations of two kinds---optical and Monte Carlo---depending on how the colliding nuclei are modeled geometrically. The optical Glauber eccentricity, based on a smooth optical-model nucleus description, can be expressed on $N_{bin}$ as~\cite{davidhq} 
\bea \label{optecc}
\epsilon_{\text{opt}} = 0.185 [\log_{10}(3\, N_{bin}/2)]^{0.96}  [\log_{10}(1136/N_{bin})]^{0.78}. 
\eea
That parametrization agrees with optical-model Glauber simulations (from which it was derived) at the percent level.
The Monte Carlo Glauber eccentricity, based on a statistical distribution of nucleons within each nucleus, differs substantially from the optical version, with elevated values especially for peripheral and central collisions as demonstrated in Ref.~\cite{gunther}. Parametrizations of the two eccentricity types are compared in Refs.~\cite{gluequad,multipoles}.

\subsection{Two-component model} \label{2comp}


The two-component model (TCM) of hadron production near mid-rapidity in high-energy nuclear collisions is based on a hypothesis of two dominant production mechanisms: (a) projectile nucleon dissociation (soft) and (b) large-angle-scattered parton fragmentation to jets (hard)~\cite{kn}. The TCM is observed to provide a consistent quantitative description of $p_t$ spectra and correlations for all collision systems at the RHIC~\cite{ppprd,porter2,porter3,hardspec,fragevo,anomalous} and LHC~\cite{alicempt,tommpt}. In more-peripheral \aa\ collisions exhibiting {\em Glauber linear superposition} (GLS) of \nn\ collisions the soft component should scale $\propto N_{part}$, and the hard component (dijet production) should scale $\propto N_{bin}$.

As noted, pair-density difference $\Delta \rho$ represents a covariance histogram. If the covariance does not change with centrality the ratio $\Delta \rho / \sqrt{\rho_{ref}}$ then scales $\propto 1/n_{ch}$. However,  $\Delta \rho$ may include contributions from several mechanisms with their own scaling behaviors. The TCM soft component of $\Delta \rho / \sqrt{\rho_{ref}}$ should scale as $N_{part}/n_{ch}$ and the hard component as $N_{bin}/n_{ch} = \nu \times N_{part} / 2n_{ch}$. That is just the scaling observed within the GLS centrality region of \auau\  collisions for soft and hard components of 2D angular correlations measured by $\Delta \rho / \sqrt{\rho_{ref}}$~\cite{anomalous}.

We can test the TCM more precisely by rescaling the fit-model amplitudes used to describe $\Delta \rho / \sqrt{\rho_{ref}}$ with the appropriate factors, either $n_{ch} / N_{part}$ (soft) or $n_{ch} / N_{bin}$ (hard). If the rescaled data are invariant on centrality over some interval we can conclude that the GLS condition does persist there, and the assumed soft- or hard-component scaling designation is correct. Rescaling results  are shown in Sec.~\ref{quadjets}.


\subsection{Azimuth quadrupole method comparisons}

The total azimuth quadrupole component of 2D angular correlations on $(\eta,\phi)$ may have contributions from several physical mechanisms or correlation features. Accurate distinctions among features and/or mechanisms is essential for correct measurement and interpretation. Two method systems can be distinguished: (a) model fits to 2D angular correlations that are not motivated a priori by physical models and (b) nongraphical numerical methods (NGNM) motivated by a specific physical model (hydrodynamic flow) that extract $v_2$ values from particle data by various numerical recipes (methods). We now summarize characteristics of the two systems.

Two-dimensional angular correlations are fitted with a simple combination of model functions determined empirically (no a priori physical assumptions). Goodness of fit is analyzed via $\chi^2$ and direct examination of 2D fit residuals (see Ref.~\cite{anomalous} for a detailed description). Physical interpretations of the model elements and significance to collision dynamics are considered only after all such phenomenological analysis is completed.

In 1D NGNM analysis one or a few sinusoids motivated by physical assumptions are fitted to 1D projections onto azimuth (the numerical procedure is algebraically equivalent to a $\chi^2$ fit with sinusoids). The goodness of fit is not evaluated, the residuals are not shown, there may remain large residuals in the 2D angular space resulting from such 1D fits~\cite{multipoles}, and the data description should then be rejected by standard analysis criteria. Nevertheless, the inferred NGNM $v_2$ data ($v_2\{EP\}$, $v_2\{2\}$, etc.) are accepted not only as adequate formal descriptions of correlation data but as {\em necessarily representing flows}, as discussed in the next section.

The amount of information extracted from data is also quite different. Projection from 2D to 1D abandons critical information that cannot then constrain the 1D fits. Unprojected 2D data {\em require} a specific choice of model elements and reject others. A 1D Fourier series can describe any 1D azimuth projection with arbitrary precision no matter what its physical origins, whereas measured 2D angular correlations from high energy nuclear collisions would reject any 1D Fourier series as the sole data model, based on standard goodness-of-fit criteria.

\section{jets vs Flows} \label{flowjet}

The present study challenges the convention assumption that a cylindrical quadrupole component of azimuth angular correlations slowly varying on $\eta$ and denoted by symbol $v_2$ represents ``elliptic flow.'' That question has emerged in a larger context where the  relative contributions of dijet production and flow phenomena to hadron yields, spectra and correlations is debated. As an alternative to the conventional flow-based narrative derived from searches for {\em a priori} defined ``signals'' of QGP formation, the role of minimum-bias dijets in yields, spectra and correlations described quantitatively by QCD calculations has emerged via several physical-model-independent analysis methods. Dijet contributions and projectile-nucleon fragmentation are accurately described by a comprehensive two-component (soft+hard) model of hadron production~\cite{kn,ppprd,hardspec,porter2,porter3,jetspec2,tomet,tommpt}. The unique TCM description had been successfully applied to all collision systems from energies near 20 GeV to LHC energies~\cite{anomalous,tommpt}.  A complementary NJ quadrupole (effectively a third component) is then observed  to be associated with only a small fraction of the  total hadron production~\cite{quadspec}. Emergence and evolution of the competing flow narrative and TCM from  the start of RHIC operations is summarized in Ref.~\cite{review}. Below we review several issues most relevant to the present study.

\subsection{Assumed necessity of flow interpretations}

In the previous section it was stated that inferred 1D NGNM $v_2$ data are accepted not only as adequate formal descriptions of correlation data but as necessarily representing flow. We observe that the statement is representative of much of the $v_2$ literature and provide here examples of highly-cited source papers. In Ref.~\cite{ollitrault} titled ``[azimuthal] Anisotropy as a signature of transverse collective flow'' only transverse flow is considered as a correlation mechanism for heavy ion collisions. In Ref.~\cite{poskvol} ``azimuthal anisotropy'' is assumed  to be synonymous with ``elliptic flow.'' The 1D azimuth distribution is represented by a Fourier series, and the Fourier coefficients $v_m$ are assumed necessarily to represent flows. The possibility of ``nonflow'' bias is acknowledged but not  related to a specific phenomenon. Reference~\cite{firstv2} assumes that $v_2$ describing all ``azimuthal anisotropy'' represents ``elliptic flow.'' The paper includes the statement ``Jets...do not contribute beyond the systematic errors for v2....'' Referenc~\cite{2004} titled ``Azimuthal anisotropy in Au+Au collisions...'' is directed toward measurement of ``elliptic flow.'' It is also broadly assumed that jets cannot contribute to hadron production below $p_t = 2$ GeV/c where flows are assumed to dominate despite strong evidence to the contrary derived from RHIC data and QCD predictions based on direct jet measurements~\cite{anomalous,hardspec,fragevo,jetspec}. 

Similar assumptions are carried over to ``higher harmonic'' flows. Reference~\cite{gunther} introduces the concept of ``triangular flow'' to account for a double-peaked structure arising from so-called ZYAM background subtraction previously interpreted in terms of ``Mach cones,'' as described in Ref.~\cite{tzyam}. In a followup study  Ref.~\cite{luzum} interprets all long-range (on $\eta$) correlations (i.e.\ ``the ridge'') as flow manifestations based on Fourier series analysis.

\subsection{Jet-related SS 2D peak vs flow interpretations}

A major difficulty for such assumed flow interpretations is the presence of a resolved same-side 2D peak at the angular origin whose systematic properties compel a jet interpretation even in more-central \aa\ collisions, although the majority of hadrons within the peak have $p_t < 2$ GeV/c~\cite{porter2,hardspec,fragevo,jetspec}. The SS 2D peak is an unanticipated component of ``azimuthal anisotropy''  that emerged from an initial study of minimum-bias 2D angular correlations carried out in the first years of RHIC operation~\cite{axialci}. The SS peak has since been fully characterized in a number of detailed followup studies~\cite{porter2,porter3,anomalous,jetspec,ytxyt}, and a jet interpretation is generally consistent with expectations from QCD~\cite{eeprd,fragevo,jetspec}.

The difficulty presented by the SS 2D peak for flow interpretations has engendered a number of conjectured alternatives, including (a) jet studies with ZYAM background subtraction, with imposed restrictive \pt\ cuts that tend to minimize and distort inferred jet azimuth structure~\cite{tzyam}, (b) ``glasma flux tubes'' coupled with radial flow as a nonjet source for the SS 2D peak, (c) initial-state (IS) \aa\ geometry fluctuations or ``lumpy initial conditions'' coupled with radial flow as a nonjet source~\cite{flowflucts,rio} and (d) interpretation of any non-$v_2$-related azimuth structure outside a small $\eta$ interval near the origin (assigned {\em by assumption} to jets) as due to ``higher harmonic'' flows~\cite{luzum}. Conjecture (a) is related to jet-biased $v_2$ data used to define the subtracted background; in effect jets are subtracted from jets~\cite{tzyam}. Conjectures (b, c) relying on radial flow as part of the peak mechanism are falsified by Refs.~\cite{hardspec,fragevo} in which all spectrum structure conventionally attributed to radial flow is found to be consistent {\em quantitatively} with pQCD predictions for jet structure, in \pp\ collisions and in \auau\ collisions for all centralities. Conjecture (d) relies on an arbitrary definition of jet structure based on \pp\ collisions that is inconsistent with \aa\ studies coupled with pQCD calculations~\cite{anomalous,fragevo,jetspec}.

\subsection{Nonjet mechanisms for the SS 2D peak}

We consider some examples of nonjet interpretations for the SS 2D peak. As with other aspects of the flow narrative attention is typically confined to more-central \aa\ collisions above the {\em sharp transition} (ST) first reported in Ref.~\cite{anomalous} where  the SS 2D peak first becomes significantly elongated on $\eta$, as first reported in Ref.~\cite{axialci}. For minimum-bias correlations (no \pt\ cuts) the SS 2D peak remains accurately described by a single 2D Gaussian on $(\eta_\Delta,\phi_\Delta)$. When typical trigger-associated \pt\ cuts are imposed the SS peak becomes non-Gaussian on $\eta_\Delta$ (develops long tails) but retains the same narrow Gaussian shape on $\phi_\Delta$ across the entire $\eta_\Delta$ acceptance. In the latter case the SS peak model is conventionally and arbitrarily split into a ``jet-like'' part narrow on $\eta_\Delta$ and an assumed ``ridge''  that is slowly-varying on $\eta_\Delta$ over a larger interval. However, the SS peak remains a monolithic 2D structure describable by a single non-Gaussian peaked model function on $\eta_\Delta$, and the observed $\eta_\Delta$-invariant SS peak azimuth width supports such a conclusion.

Alternative mechanisms for  nonjet SS ``ridge'' formation are based on conjectured initial-state (IS) geometry deviations from a smooth ``almond shaped'' \aa\ overlap region in configuration space {\em coupled with strong radial flow} to produce specific final-state momentum azimuth correlations. The non-smooth IS geometry structure is attributed to (a) random fluctuations in the IS nucleon distributions within nuclei~\cite{gunther}, (b) glasma flux tubes~\cite{mosch1,mosch2}, (c) IS geometry fluctuations~\cite{flowflucts} and (d) ``lumpy initial conditions''~\cite{rio} as examples. Monte Carlo models for (c,d) include AMPT (with final-state rescattering)~\cite{ampt} and NexSpheRio (with hydro expansion)~\cite{rio}.

\subsection{Evidence against  nonjet SS peak mechanisms}

Challenges  to such conjectures  were provided in previously published papers and are not the subject of the present study. However, we briefly summarize a few problematic issues for nonjet SS 2D peak mechanisms.

For any proposed IS geometry structure in configuration space to produce observable azimuth structure in final-state momentum space requires strong transverse flow (generated by multiple parton and/or hadron rescattering) as a radial transport mechanism. However, the spectrum structure conventionally interpreted to represent radial flow (via an assumed blast-wave spectrum model) has been quantitatively described in  terms of pQCD jets~\cite{hardspec,fragevo,jetspec}. Radial flow is thereby excluded as a significant collision mechanism. As noted, radial flow should be generated by multiple rescattering of partons and/or hadrons during collision evolution. However, the continuing presence of resolved jet-related structure with the expected \pt\ structure corresponding to 3 GeV jets contradicts the presence of significant rescattering~\cite{jetspec}.

The case against glasma flux tubes has been presented in Refs.~\cite{glasma,glasma2}. The lack of radial flow immediately eliminates such a mechanism even assuming glasma flux tubes might play some role. If radial flow did exist it must vary strongly with $\eta_\Delta$ to produce  the observed MB SS 2D peak structure (with large curvature on $\eta_\Delta$). However, elliptic flow is understood to be a modulation of the same radial flow, and $v_2$ is nearly uniform on $\eta$ within  the STAR TPC acceptance. Thus, radial flow must assume two contradictory longitudinal structures within the conventional flow narrative which is impossible. The glasma flux-tube mechanism is also eliminated by comparison with observed \pt\ correlation structure described below~\cite{glasma2}. 

The centrality dependence of correlation structure reported in Refs.~\cite{anomalous,davidhq} also contradicts nonjet SS 2D peak mechanisms. Below the sharp transition the SS 2D peak is narrow on $\eta_\Delta$ as well as $\phi_\Delta$ and fully resolved as an isolated 2D peak within the STAR TPC. Its properties are consistent with expected intrajet correlations. The AS dipole is in turn consistent with interjet (back-to-back jet) correlations. The proposed nonjet SS 2D peak mechanisms cannot reproduce such a localized peak structure and are thus falsified by  the data. Whereas the SS 2D peak and AS dipole centrality dependences are closely coupled for all \aa\ centralities, as expected for dijets,  the NJ quadrupole has a unique centrality trend very different from the jet-related structure~\cite{anomalous,davidhq}, the main subject of  the present study. Below  the ST there is no significant evidence for IS geometry fluctuations in spectra or correlations. All data are consistent with the reference TCM corresponding to Glauber linear superposition (GLS) of \nn\ collisions and a smooth IS geometry. With increasing centrality above the ST there is no discontinuity in the properties of correlation structure, only a rapid change in the {\em rate of variation} of jet-related properties near the ST and no change in the universal NJ quadrupole trend. Thus, we find no point on centrality where one might conclude that jet structure has disappeared and flows increase to dominate collision dynamics.


It has become fashionable to project 2D angular correlations onto 1D azimuth (possibly with some restrictions on pair $\eta$ acceptance as well as imposed \pt\ cuts) and represent the result by a Fourier series~\cite{gunther,luzum}. The various Fourier coefficients are  then interpreted as representing flows (elliptic, triangular, etc.). In effect the first term in the model of Eq.~(\ref{eq1}) is replaced by extra Fourier series elements in the second term. We have challenged such procedures in Refs.~\cite{multipoles,sextupole} wherein we show that some Fourier terms may then represent multiple distinct correlation features, and the 1D Fourier series cannot possibly represent 2D angular correlations, must be rejected by standard fit criteria. The changes in Fourier amplitudes generated by the ``higher harmonic'' approach always sum to a narrow SS 1D Gaussian that serves as an approximation to the elongated tails on the SS 2D peak~\cite{sextupole}. And below the ST the 1D Fourier series without SS 2D peak model obviously fails catastrophically.







The \pt\ structure of angular correlations has been studied extensively. In Ref.~\cite{ytxyt}  symmetrized combinatoric transverse rapidity correlations on  $y_t \times y_t$ in \auau\ collisions for all centralities were found to have the same TCM structure as those for \pp\ collisions reported in Refs.~\cite{porter2,porter3}. The $y_t \times y_t$ hard component corresponds to jet-like angular correlations and persists even in central collisions, albeit with quantitative modifications. In Refs.~\cite{ptscale,ptedep} mean-\pt\ fluctuations were inverted to recover the underlying \pt\ angular correlations which are consistent with jet structure for \pp\ collisions and for \auau\ collisions of all centralities. The $\eta$-elongated parts of the SS 2D peak retain the \pt\ structure of jets. Flow mechanisms cannot generate the observed \pt\ correlation structure with typical values near 1 GeV/c. In Ref.~\cite{tommpt} the mean-\pt\ data from Ref.~\cite{alicempt} were described accurately by a TCM including dijet production as a principal mechanism along with a universal and fixed soft component. For all collision systems studied at the LHC the only significant source of mean-\pt\ {\em variation} is dijet production.

\subsection{Summary}

From arguments summarized above and other evidence we conclude that (a) a separate 2D model for the SS peak is required in all cases, which may include a non-Gaussian shape on $\eta_\Delta$ for more-central \aa\ collisions, (b) the SS 2D peak is dominated by dijets, (c) the SS 2D peak is closely related to a separate AS 1D peak corresponding to back-to-back jets in all cases and (d) a distinct NJ quadrupole is not directly related to the SS 2D peak or AS 1D peak and demonstrates independent systematic trends. We now turn to the main purpose of this study, a consideration of the ``elliptic flow'' interpretation for the NJ quadrupole in relation to alternative interpretations.


\section{2D angular autocorrelations} \label{angcorr}

We describe data volumes, example 2D data histograms, principal features of the angular correlations and fitting procedures used to derive jet-related and nonjet quadrupole energy and centrality systematics.

\subsection{Data histograms}

The analyses reported in Refs.~\cite{davidhq,anomalous} were based on 6.7M and 1.2M Au-Au collisions at $\sqrt{s_{NN}} =$ 62.4 (year 2004) and 200 GeV (year 2001) respectively, observed with the STAR TPC. The momentum acceptance was defined by transverse momentum $p_t > 0.15$ GeV/c, pseudorapidity $|\eta| < 1$ and $2\pi$ azimuth. Au-Au collision centrality was defined as in Ref.~\cite{centmeth}. Minimum-bias event samples were divided into 11 centrality bins: nine $\sim10$\% bins from 100\% to 10\%, the last 10\% divided into two 5\% bins. The corrected centrality of each bin as modified by tracking and event-vertex inefficiencies was determined 
with a running-integral procedure. Centralities from \mbox{N-N} collisions to central \auau\  were thereby determined to about 2\% uncertainty.

 \begin{figure}[h]
  \includegraphics[width=1.65in,height=1.58in]{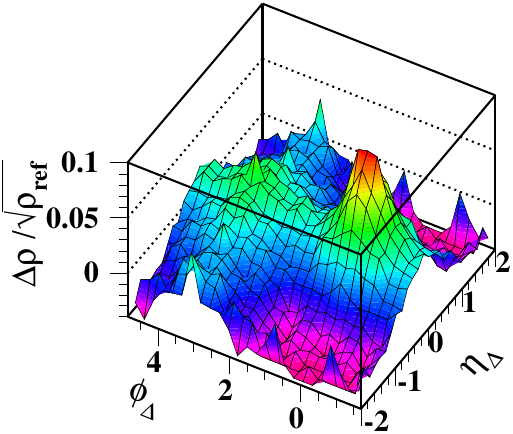}
 \put(-90,99) {\bf (a)}
\includegraphics[width=1.65in,height=1.58in]{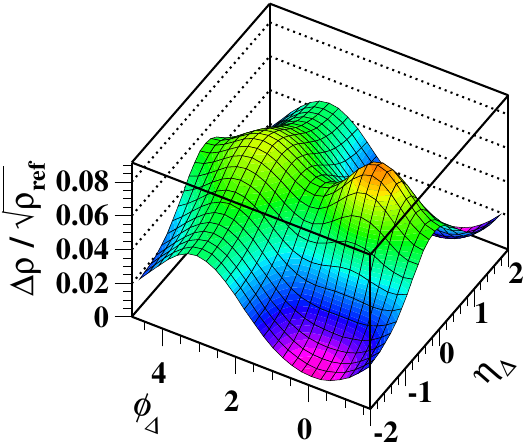}
  \put(-90,99) {\bf (b)} \\
\includegraphics[width=1.65in,height=1.58in]{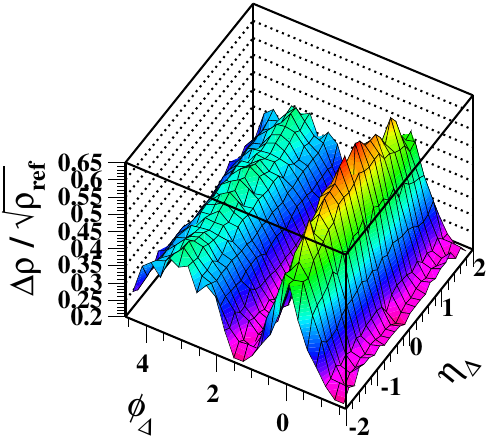}
 \put(-90,99) {\bf (c)}
\includegraphics[width=1.65in,height=1.58in]{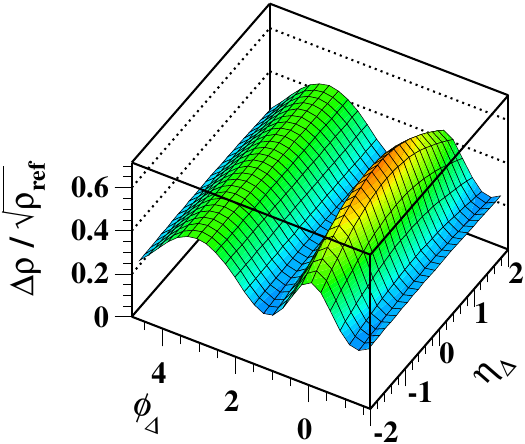}
 \put(-90,99) {\bf (d)}
 \caption{\label{fig1}
(Color online) Left: 
2D angular autocorrelations for 200 GeV \auau\  collisions and (a) 83-93\% centrality ($\sim$N-N collisions) and (c) 0-5\% centrality. Histograms from 62 GeV collisions have the same general features with quantitative differences. 
Right: Two-dimensional model fits to histograms in the left panels without the BEC-electron component. 
 } 
 \end{figure}

Figure~\ref{fig1} (left panels) shows 200 GeV 2D angular correlations for (corrected) 83-93\% ($\approx$ N-N collisions) and 0-5\% centrality bins. Angular correlations for 62 GeV have similar features but with quantitative differences.
Within the STAR TPC acceptance the minimum-bias correlation data from \auau\  collisions include three principal components: (a) a same-side (SS) 2D peak at the origin on $(\eta_\Delta,\phi_\Delta)$ well approximated by a 2D Gaussian for all minimum-bias data, (b) an away-side (AS) 1D peak on azimuth or ``ridge'' well approximated by AS azimuth dipole $[1 - \cos(\phi_\Delta)]/2$ for all minimum-bias data and uniform to a few percent on $\eta_\Delta$ (having negligible curvature), and (c) an azimuth quadrupole $\cos(2\phi_\Delta)$ also uniform on $\eta_\Delta$ to a few percent over the full angular acceptance of the STAR TPC. Other components consist of a sharp 2D exponential peak at (0,0) and a narrow 1D peak on $\eta_\Delta$. That phenomenological description does not rely on any physical interpretation of the components. 

Based on subsequent comparisons of observed data systematics with theory the components (a) and (b) together have been interpreted to represent minimum-bias jets or minijets~\cite{fragevo,anomalous}. Component (c), identified as the nonjet azimuth quadrupole, has been conventionally attributed to elliptic flow~\cite{2004}.  However, alternative mechanisms have been proposed~\cite{gluequad,boris}. The 2D exponential represents Bose-Einstein correlations and electron pairs from photoconversions, and the narrow 1D peak on $\eta_\Delta$ is attributed to projectile-nucleon dissociation. Reinterpretation of the NJ quadrupole based on comparison with jet-related systematics is the main subject of this study.

\subsection{Two-dimensional fit model} \label{fitmodel}

In this study we emphasize correlation components (a), (b) and (c). The corresponding 2D model function is~\cite{axialci,anomalous,davidhq}
\bea \label{estructfit}
\frac{\Delta \rho}{\sqrt{\rho_{ref}}} \hspace{-.02in}  & = &  \hspace{-.02in}
A_0+  A_{2D} \, \exp \left\{- \frac{1}{2} \left[ \left( \frac{\phi_{\Delta}}{ \sigma_{\phi_{\Delta}}} \right)^2 \hspace{-.05in}  + \left( \frac{\eta_{\Delta}}{ \sigma_{\eta_{\Delta}}} \right)^2 \right] \right\} \nonumber \\
&+&  \hspace{-.02in} A_{D}\, \{1 +\cos(\phi_\Delta - \pi)\} / 2 + \hspace{-.02in}  A_{Q}\, 2\cos(2\, \phi_\Delta).
\eea
A 1D Gaussian on $\eta_\Delta$ (soft component, negligible in more-central \auau\  collisions) and 2D exponential (very narrow in more-central \auau\  collisions) are omitted from Eq.~(\ref{estructfit}) for simplicity but were included in the analyses of Refs.~\cite{davidhq,anomalous}. Equation~(\ref{estructfit}) is a more-detailed version of Eq.~(\ref{eq1}).

Nonjet quadrupole measure $A_Q$ as defined by Eq.~(\ref{estructfit}) is statistically compatible with jet-related measures $A_{2D}$ and $A_D$ (i.e.\ all are per-particle measures), permitting quantitative comparisons between jet-related and nonjet quadrupole systematics.  The quadrupole amplitude is related to conventional measure $v_2\{2D\}$ by $A_{\rm Q}\{2D\} = \rho_0(b)\, v^2_2\{{\rm 2D}\}$, where $\rho_0(b) = dn_{\rm ch} / 2\pi d\eta$ is the single-particle 2D angular density, and symbol $\{{\rm 2D}\}$ denotes parameters  inferred from 2D model fits to angular correlations as described in Refs.~\cite{flowmeth,gluequad,davidhq,anomalous}.

\section{Two-dimensional Model Fits} \label{2dfit}

Figure~\ref{fig1} (right panels) shows typical 2D model fits compared to corresponding data histograms in the left panels. For each data histogram fits are initiated from many different combinations of initial starting parameters (typically 100-1000) to insure achievement of global $\chi^2$ minima. The fit residuals are typically consistent with bin-wise statistical uncertainties. The general evolution with centrality is monotonic increase of the SS 2D peak and AS dipole amplitudes (jet-related structures), substantial increase of the SS peak $\eta_\Delta$ width, rapid decrease to zero of the 1D Gaussian on $\eta_\Delta$~\cite{axialci,ptscale,anomalous} and non-monotonic variation of the nonjet quadrupole~\cite{davidhq}.

\subsection{Jet-related structures} \label{jetprops}

\begin{figure}[h]
   \includegraphics[width=1.65in,height=1.63in]{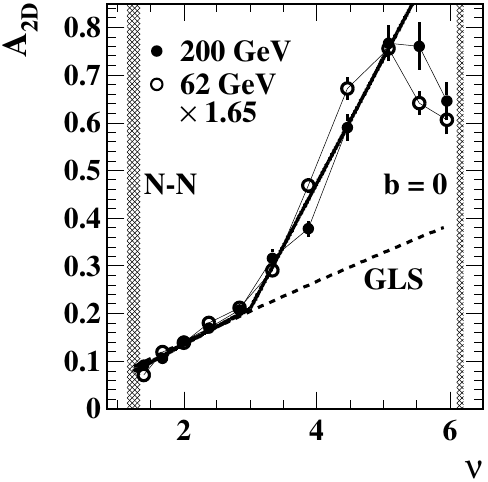}
  \put(-30,25) {\bf (a)}
  \includegraphics[width=1.65in,height=1.66in]{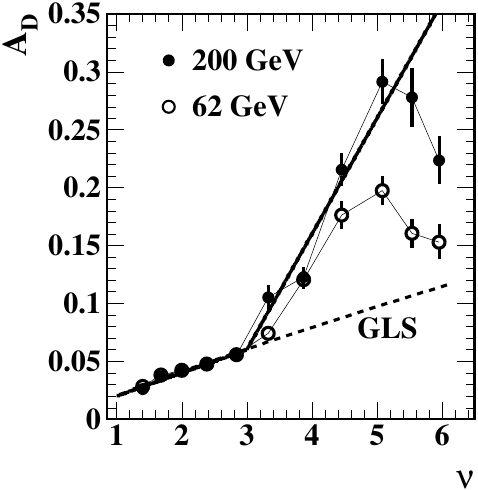}
  \put(-30,25) {\bf (b)} \\
   \includegraphics[width=1.65in,height=1.66in]{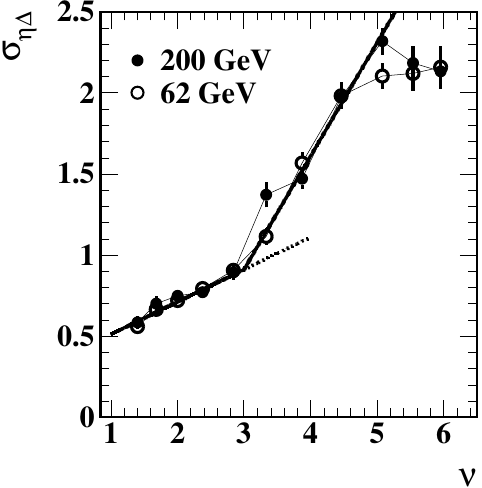}
  \put(-30,25) {\bf (c)}
   \includegraphics[width=1.65in,height=1.66in]{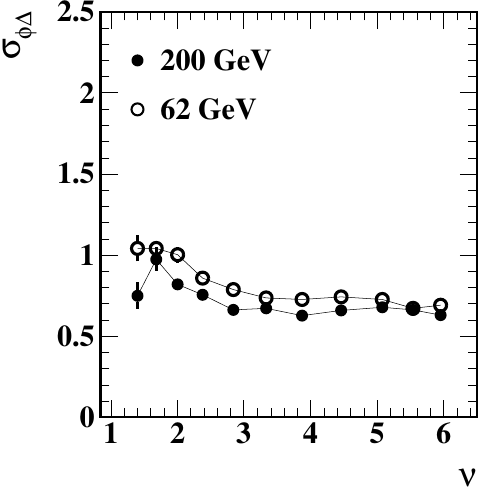}
  \put(-30,25) {\bf (d)}
\caption{\label{jetfits}
Centrality dependence of fit parameters from Eq.~(\ref{estructfit}) for (a) same-side (SS) 2D peak amplitude, (b) away-side (AS) 1D peak amplitude, (c) SS peak $\eta$ width, (d) SS peak $\phi$ width.
 } 
 \end{figure}

Figure~\ref{jetfits} shows fit results for jet-related structures from Ref.~\cite{anomalous} where they are extensively discussed. The dashed curves in the upper panels indicate a Glauber linear superposition (GLS) trend expected for transparent \aa\ collisions. The jet-related amplitudes follow that trend from \nn\ collisions to a {\em sharp transition} at $\nu \approx 3$ corresponding to $\sigma/\sigma_0 \approx 50$\%. Above that point the amplitudes increase relative to the GLS trend in a manner consistent with a modification of parton fragmentation that conserves parton energy within resolved jets~\cite{fragevo,jetspec}.

Figure~\ref{jetfits} (lower panels) show the $\eta$ and $\phi$ widths of the SS 2D peak. Strong elongation on $\eta$ of the SS peak in more-central \auau\  collisions was first reported in Ref.~\cite{axialci}. The physical mechanism for elongation is currently intensely debated, as discussed in Sec.~\ref{flowjet} and for example Refs.~\cite{glasma,glasma2}. It is notable that the SS peak azimuth width actually {\em decreases} with increasing centrality and is the same for all $\eta_\Delta$ values within the TPC acceptance, implying a single monolithic SS 2D peak. Conjectured mechanisms for jet modification and/or parton energy loss that rely on multiple scattering and/or gluon bremsstrahlung must confront that decrease.

\subsection{Nonjet quadrupole}

\begin{figure}[h]
   \includegraphics[width=1.65in,height=1.66in]{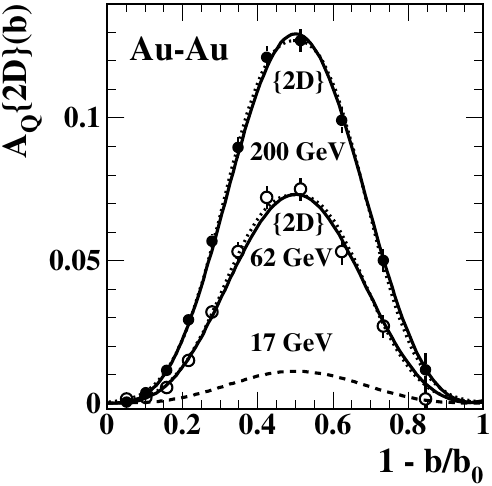}
  \put(-25,85) {\bf (a)}
  \includegraphics[width=1.65in,height=1.66in]{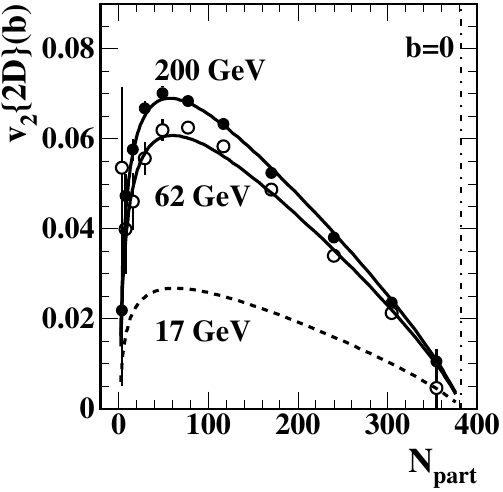}
  \put(-30,85) {\bf (b)}
\caption{\label{quadfits}
(a) Nonjet quadrupole amplitude $A_Q\{2D\}(b)$ plotted on relative impact parameter $b/b_0 = \sqrt{\sigma / \sigma_0}$ with $b_0 \approx 14.7$ fm.
(b) Corresponding values for $v_2\{2D\}(b)$ on participant number $N_{part}$. Solid and dashed curves are defined by Eq.~(\ref{loglog}).
 } 
 \end{figure}

Figure~\ref{quadfits} summarizes 2D fit results for $A_Q\{2D\}(b)$  (left panel) and corresponding values of $v_2\{2D\}(b)$ (right panel) for comparison with published $v_2$ measurements. The left panel shows fit results for 200 GeV (solid dots) and 62 GeV (open circles) data: strong increase to mid-central collisions followed by reduction to zero for central collisions. In the right panel the $v_2$ trend suggests substantial ``elliptic flow'' for the most-peripheral centrality bin approximating N-N collisions. The solid and dashed curves are defined in the next subsection. The dashed curves correspond to NA49 $v_2\{\text{EP}\}$ measurements at 17 GeV (see Sec.~\ref{compare}) that provide a reference for inferred energy-dependence systematics. 

The plotting format in the left panel reveals an interesting centrality trend common to all energies between 13.5 GeV and 200 GeV: $A_Q(b)$ data approximate a Gaussian trend on relative impact parameter $b/b_0$ ($b_0 \approx 14.7$ fm for \auau\ collisions). The dotted curves just visible behind the solid curves are modified Gaussians symmetric about the midpoint.\footnote{The functional form is $\exp\{-|(x - 0.5)/0.178|^{2.4}/2\}$.}

The plotting format in the right panel is the more conventional $v_2(b)$ vs $N_{part}(b)$. The choice of both the correlation measure and centrality measure can be questioned. Quantity $v_2$ is the square root of per-pair measure $v_2^2$ that is actually inferred from the data pair ratio $\Delta \rho / \rho_{ref}$ and therefore tends to deemphasize systematic variations. 

Centrality measure $N_{part}$ biases the visual presentation to favor the more-central 50\% of the \auau\ fractional cross section $\sigma / \sigma_0$ (lying above $N_{part} \approx 50$) and deemphasizes the more-peripheral half of the total cross section. Thus, the relation between \pp\ collisions as a reference and novel behavior in more-central \auau\ collisions is obscured.

As demonstrated below, substantial changes in jet systematics occur near  $N_{part} \approx 50$ (deviations from the GLS trend above that point)~\cite{anomalous}. The lower half of the fractional cross section provides an essential GLS reference and should remain visually accessible, but the GLS trend for more-peripheral collisions is effectively concealed by the $N_{part}$ centrality measure. Proper interpretation of nonjet quadrupole data {\em in relation to jet data} is then compromised.  Alternative centrality measures include $b/b_0$ (left panel), fractional cross section $\sigma/\sigma_0$ and mean participant path length $\nu$.

\subsection{Combining energy and centrality trends}

The data in Fig.~\ref{quadfits} (left panel) reveal two interesting features: (a) Data for all energies above 13 GeV are described by the same centrality variation (solid and dashed curves), and (b) quadrupole amplitude $A_Q$ scales with energy approximately as $\log(\sqrt{s_{NN}})$. A similar energy scaling was observed for per-particle $\langle p_t \rangle$ fluctuations/correlations attributed to minijets~\cite{ptedep}. An algebraic model describing quadrupole energy and centrality trends (solid and dashed curves) is now derived.

\begin{figure}[h]
  \includegraphics[width=1.65in,height=1.635in]{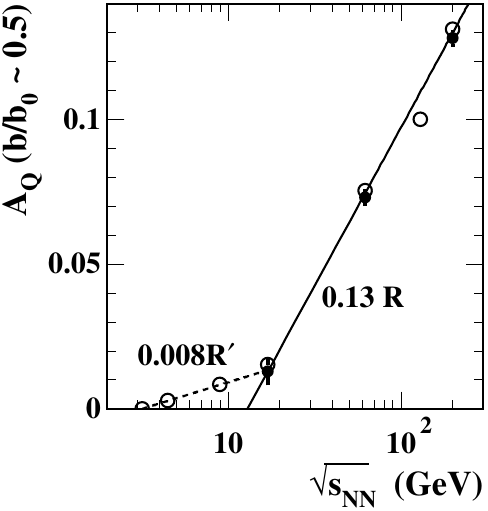}
   \put(-80,75) {\bf (a)}
 \includegraphics[width=1.65in,height=1.66in]{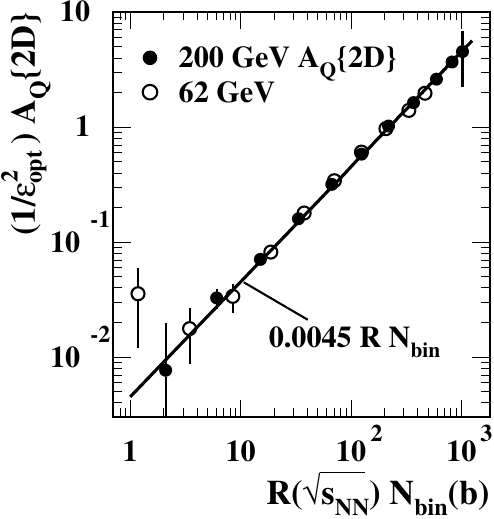}
  \put(-80,75) {\bf (b)}
\caption{\label{energycent}
(a) The energy dependence of azimuth quadrupole amplitude $A_Q = \rho_0 v_2^2$ evaluated at the maximum of that parameter on centrality. The trend $0.13 R(\sqrt{s_{NN}})$ (solid line) defined above is included in Eq.~(\ref{loglog}).
(b)  Nonjet quadrupole data (points) for 62 and 200 GeV \auau\ compared to the trend defined by Eq.~(\ref{loglog}) (straight line).
 } 
 \end{figure}

Figure~\ref{energycent} (left panel) shows the trend on collision energy of $A_Q(b)$ maximum values at $b/b_0 \approx 0.5$ (points to the right of 10 GeV). Published $v_2$ measurements have been converted to $A_Q$ values based on corresponding multiplicity densities and the relation $A_Q = \rho_0 v_2^2$. 
We observe two energy regimes: Below 13.5 GeV (Bevalac-AGS) the $A_Q$ values are small and evolve with energy from negative to positive in response to the kinematic influence of spectator nucleons. The actual energy trend for {\em collective expansion} of participants is not known.
The dashed line is defined by energy scaling factor $R'(\sqrt{s_{NN}}) \equiv \ln(\sqrt{s_{NN}} / \text{3.2 GeV})$ with coefficient 0.008. Above 13.5 GeV (SPS-RHIC) the rate of increase becomes dramatically larger. 
The solid line is defined by scaling factor $R(\sqrt{s_{NN}}) \equiv \ln\{\sqrt{s_{NN}}/13.5~ \text{GeV}\} /\ln(200/13.5)$ with $A_Q$ intercept $13.5\pm 0.5$ GeV and coefficient $0.13 \pm 0.01$.

Figure~\ref{energycent} (right panel) shows $A_Q$ data  plotted in the form $(1/\epsilon_\text{opt}^2) A_Q\{2D\}(b)$ {\em vs}\, $R(\sqrt{s_{NN}})\, N_{bin}(b)$, where $N_{bin}$ is the number of binary N-N collisions and $\epsilon_{opt}$ is the eccentricity derived from an {\em optical}-Glauber Monte Carlo simulation (Sec.~\ref{centecc}). We observe empirically that for all \auau\ collisions above 13 GeV the $A_Q\{2D\}$ data are described accurately by the relation (solid line)
\bea \label{loglog}
 A_Q\{2D\}(b,\sqrt{s_{NN}}) \hspace{-.04in} &=& \hspace{-.04in} C_0\, R(\sqrt{s_{NN}})\, N_{bin}(b)\,\epsilon_\text{opt}^2(b),
\eea
with coefficient $C_0$ defined by $1000 C_0 = 4.5\pm 0.2$. 
Equation~(\ref{loglog}) accurately describes the measured $p_t$-integrated nonjet azimuth quadrupole in \auau\ collisions for all centralities down to N-N collisions and all energies down to $\sqrt{s_{NN}} \approx $13 GeV (as in Fig.~\ref{energycent} -- left). It defines the solid and dashed curves in Fig.~\ref{quadfits}. 

\subsection{Comparisons with other methods} \label{compare}

NGNM $v_2$ measurements are conventionally interpreted to represent some combination of ``elliptic flow'' and ``nonflow''~\cite{poskvol} with several proposed sources for the latter such as resonances, Bose-Einstein correlations and jets~\cite{2004}. Separating sinusoids attributed to ``flow'' from other correlation structure is a long-standing problem not resolved by NGNM analysis. Various strategies have been proposed to reduce ``nonflow,'' including cuts on $\eta$ to exclude an interval on $\eta_\Delta$ near the origin. The most common methods and their biases have been compared to 2D model fits on $(\eta_\Delta,\phi_\Delta)$~\cite{flowmeth,gluequad,tzyam,multipoles,sextupole}.

\begin{figure}[h]
  \includegraphics[width=1.65in,height=1.635in]{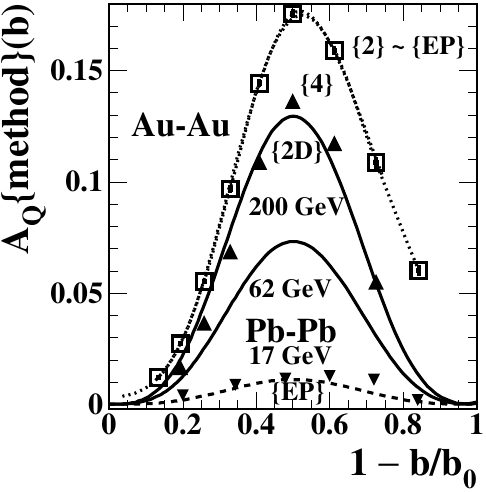}
  \put(-85,65) {\bf (a)}
 \includegraphics[width=1.65in,height=1.635in]{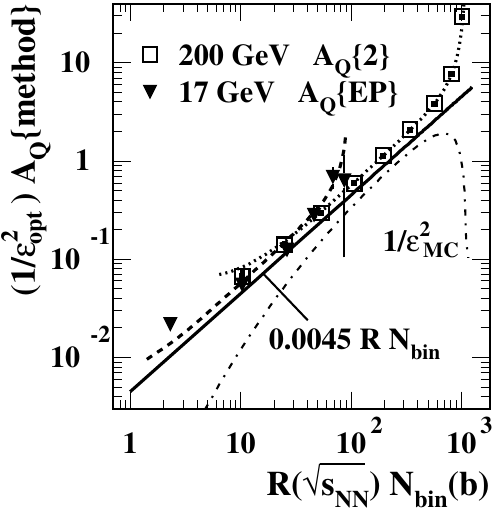}
  \put(-85,65) {\bf (b)}
\caption{\label{v2ep}
(a) Quadrupole results from 2D model fits presented in Fig.~\ref{quadfits} (solid and dashed curves) compared to published $v_2$ measurements obtained with NGNM methods (points). The dotted curve includes a jet-related contribution obtained from the measured SS 2D peak systematics in Fig.~\ref{jetfits}.
(b)  2D model-fit trend from Eq.~(\ref{loglog}) (solid line) compared to published $v_2$ measurements from NGNM methods (points). The dotted and dashed curves include a calculated contribution from the measured SS 2D peak. 
 } 
 \end{figure}

Figure~\ref{v2ep} (left panel) shows multiparticle cumulant measurements $A_Q\{2\}$ (open squares) and $A_Q\{4\}$ (solid upright triangles) from 200 GeV \auau\ collisions~\cite{2004} and event-plane measurements $A_Q\{EP\}$ (solid inverted triangles) from 17 GeV \pbpb\ collisions~\cite{na49} compared with Eq.~(\ref{loglog}) (solid and dashed curves). In Ref.~\cite{flowmeth} it was demonstrated that $v_2\{2\} \approx v_2\{EP\}$ to within 5\% (consistent with experiment~\cite{2004}). ``Nonflow'' contributions to those methods are discussed in the Appendix.  Published uncertainties for the $A_Q\{2\}$ measurements multiplied by factor 10 to make them visible are shown as bars within the open squares.
It is claimed that $v_2\{4\}$ eliminates ``nonflow'' arising from small clusters of particles (e.g., jets)~\cite{borg,starv24}, and those measurements are indeed closer to the $A_Q\{2D\}$ trend (solid curve), but there are still quite significant differences. The NA49 $v_2\{\text{EP}\}$ measurements (inverted triangles)~\cite{na49} provide a reference for the energy-dependence systematics. 

Figure~\ref{v2ep} (right panel) shows the $A_Q\{2\}$ and $A_Q\{EP\}$ measurements compared with Eq.~(\ref{loglog}) (solid line). Deviations of the NGNM measurements from the linear trend are consistent with expected bias contributions from jet structure (SS 2D peak)~\cite{gluequad} as described  in the Appendix. The dotted curve passing through the 200 GeV measurements is a combination of Eq.~(\ref{loglog}) with jet-related contribution $A_Q\{SS\}$ derived from jet-related correlation properties presented in Sec.~\ref{jetprops}. The same procedure generated the dotted curve in the left panel. The dashed curve approximating the 17 GeV measurements includes the same jet-related $A_Q\{SS\}$ contribution scaled down with energy according to $R(\sqrt{s_{NN}})$ from Eq.~(\ref{loglog}).
The dash-dotted curve indicates what the $A_Q\{2D\}$ data trend (solid line) would be if $\epsilon_{opt}$ from Eq.~(\ref{optecc}) were replaced by Monte Carlo $\epsilon_{MC}$ adopted from Ref.~\cite{epsnpart} (and see Ref.~\cite{multipoles}). The Monte Carlo eccentricity trend may compensate partially for the jet contribution to $A_Q\{2\}$.

\section{quadrupole vs jet trends} \label{quadjets}

A unique finding of Ref.~\cite{anomalous} was the ``sharp transition'' (in jet structure trends from \auau\ collisions) separating Glauber linear superposition and apparent \aa\ transparency within the more-peripheral half of the total cross section and strong deviations from the GLS trend still consistent with a pQCD description within the more-central half~\cite{hardspec,fragevo}. The NJ quadrupole data reported in Ref.~\cite{davidhq} demonstrated remarkably simple energy and centrality trends for all \auau\ (or Pb-Pb) centralities and all energies above 13 GeV. We make direct comparisons between jet-related systematics and the nonjet quadrupole. We emphasize the relation of data trends to initial-state geometry parameters and possible QCD mechanisms. 

\begin{figure}[h]
  \includegraphics[width=1.65in,height=1.66in]{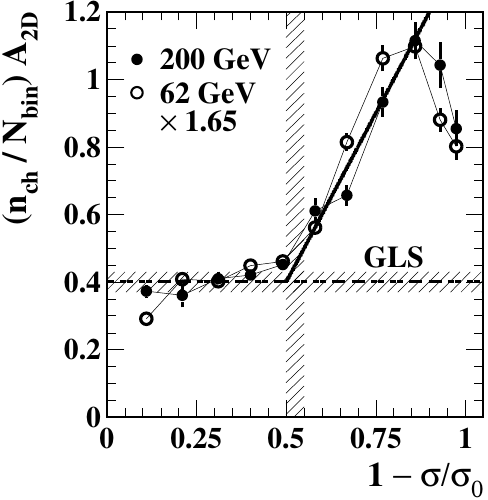}
   \put(-85,70) {\bf (a)}
 \includegraphics[width=1.65in,height=1.66in]{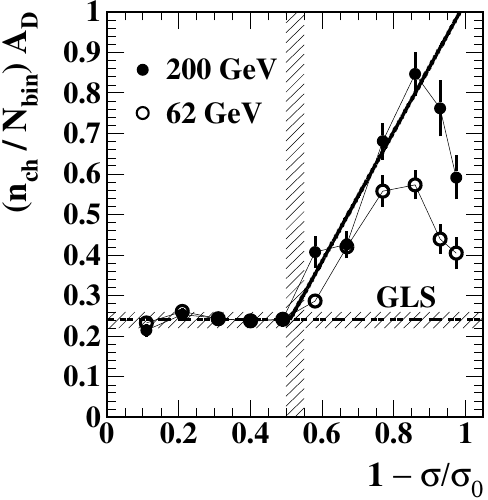}
  \put(-85,70) {\bf (b)}
\caption{\label{mike}
(a)  Same-side 2D peak amplitude vs fractional cross section. The amplitude has been rescaled to compare the covariance density $\Delta \rho$ to the number of binary \nn\ collisions.
(b) Away-side 1D peak amplitude rescaled in the same way. The constant trends are consistent with pQCD dijet production within transparent \aa\ collisions. The hatched bands mark a {\em sharp transition} (ST) in jet-related systematics.
 } 
 \end{figure}

\subsection{Minimum-bias jet systematics}

Figure~\ref{mike} shows jet-related SS 2D peak $A_{2D}$ (2D Gaussian) and AS 1D peak $A_D$ (dipole) amplitudes vs centrality measured by fractional cross section. The hatched bands show the position of the sharp transition (ST) near 50\% ($\nu \approx 3$ or $N_{part} \approx 50$). The per-particle peak amplitudes $A_X$ are rescaled by factor $n_{ch} / N_{bin}$ because those hard-component amplitudes are expected to scale with $N_{bin}$ (as described in Sec.~\ref{2comp}). The covariance in the numerator of $\Delta \rho / \sqrt{\rho_{ref}}$ is then compared directly with the number of initial-state \nn\ binary collisions rather than the number of final-state hadrons. We observe that below the ST the ``jet-related'' amplitudes are systematically consistent with a constant value ($N_{bin}$ scaling from \pp\ collisions) as expected for dijet production in a transparent system, thereby buttressing the jet interpretation. Above the ST the amplitudes increase substantially relative to the GLS trend, but the corresponding changes within \pt\ spectra are still described quantitatively within a pQCD context~\cite{fragevo}.

In the left panel the SS peak amplitudes for 62 GeV rescaled by factor $1/R(\text{62 GeV}) = 1.65$ (with dijet intercept at 10 GeV consistent with Ref.~\cite{jetspec2}) are then equivalent to the 200 GeV amplitudes. In  the right panel the unrescaled AS peak amplitudes for the two energies agree closely within the GLS interval. For in-vacuum dijets we expect the SS peak amplitude (sum of individual jets projected onto 1D $y_z$) to exhibit a $\log(\sqrt{s_{NN}})$ trend due to the increase of the kinematically-allowed longitudinal rapidity $y_z$ interval, whereas the AS peak amplitude [representing the dijet density on 2D $(y_{z1},y_{z2})$] should increase more slowly or not at all with energy~\cite{jetspec,ppcms}. We observe just such trends within the transparency interval. Thus, comparison of SS and AS centrality and energy trends strongly supports a dijet interpretation for those correlation structures, but also reveals a significant quantitative change in some jet-related correlation properties above the ST. 
The substantial increase of jet-related amplitudes (and SS $\eta$ width) above the ST corresponds quantitatively to possible changes in parton fragmentation that still conserve the full parton energy within the resolved jet structure~\cite{bw}. That description is supported by spectrum analysis~\cite{hardspec,fragevo} and correlation analysis~\cite{jetspec}.

\subsection{Nonjet quadrupole systematics}

\begin{figure}[h]
  \includegraphics[width=1.65in,height=1.635in]{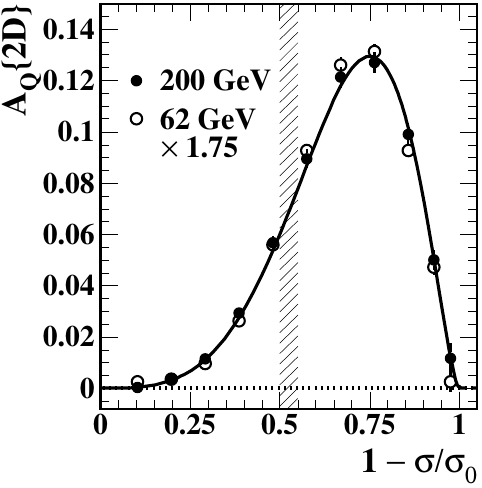}
   \put(-30,30) {\bf (a)}
 \includegraphics[width=1.65in,height=1.66in]{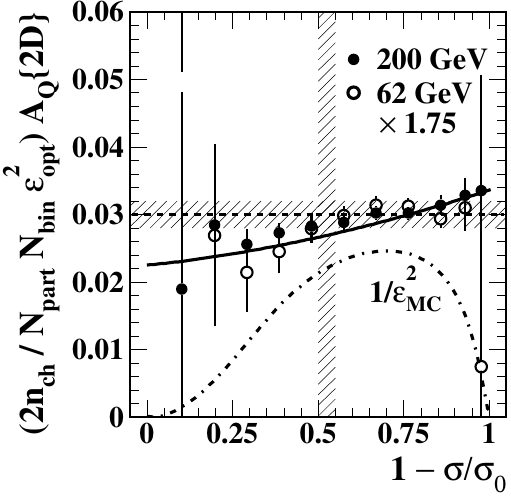}
   \put(-30,30) {\bf (b)}
\caption{\label{quadcomp}
(a) Azimuth quadrupole amplitudes vs fractional cross section. The 62 GeV data are scaled up according to factor $1/R(\sqrt{s_{NN}}) = 1.75$ and coincide at the percent level with the 200 GeV data. The curve is Eq.~(\ref{loglog}). The hatched band indicates that the nonjet quadrupole achieves 60\% of its maximum amplitude within transparent \auau\ collisions.
(b)  $A_Q$ data rescaled to participant nucleon pairs and further rescaled according to Eq.~(\ref{loglog}). Both the rescaled and unrescaled data are systematically consistent with Eq.~(\ref{loglog}). The dash-dotted curve shows the data trend that would result from rescaling with $\epsilon_{MC}$ instead of $\epsilon_{opt}$.
 } 
 \end{figure}

Figure~\ref{quadcomp} (left panel) shows quadrupole amplitude $A_Q\{2D\}$ with the 62 GeV data rescaled by factor $1/R(\text{62 GeV}) = \ln(200/13.5) / \ln(62/13.5) = 1.75$. The close overall agreement is consistent with Eq.~(\ref{loglog}). The point-to-point agreement demonstrates the accuracy of the analysis method, with deviations at the few-percent level for two distinct data volumes. The hatched band represents the sharp transition in jet properties. It is remarkable that in the more-peripheral centrality interval, where 3 GeV partons appear as in-vacuum jets (hadron $\langle p_t \rangle \approx 1$ GeV/c) with no modification and we describe \aa\ collisions as {\em transparent}, the nonjet quadrupole conventionally interpreted to represent elliptic flow of a dense, strongly-interacting QGP increases to 60\% of the maximum value as measured by $A_Q\{2D\}$ or the maximum value as measured by $v_2\{2D\}$ in Fig.~\ref{quadfits} (right).

Figure~\ref{quadcomp} (right panel) shows a variant of Fig.~\ref{energycent} (right panel) in which the $A_Q$ data are first rescaled by factor $2n_{ch} / N_{part}$ ($n_{ch} A_Q \propto$ number of quadrupole-correlated pairs) to compare the quadrupole covariance in the numerator to initial-state participant pairs rather than final-state hadrons and then by factor $1/N_{bin} \epsilon_{opt}^2$ corresponding to Eq.~(\ref{loglog}). The data are systematically consistent with a constant value (dashed line). The solid curve represents the same rescaling applied to the Eq.~(\ref{loglog}) expression for $A_Q$ and is also consistent with the data. 

Figure~\ref{quadfits} (right panel) shows $v_2\{2D\} = 0.022$ for the most peripheral 200 GeV centrality bin (approximately \nn\ collisions). In Fig.~\ref{quadcomp} that \nn\ value is consistent with a simple scaling trend describing all \auau\ centralities. It is also notable that $v_2\{2D\}$ in the \nn\ limit of \auau\ collisions is consistent with a pQCD color-dipole {\em prediction} $v_2 \approx 0.02$ for pions from \pp\ collisions~\cite{boris}. 

Collision eccentricity can also be modeled by a {\em participant-nucleon} or {\em Monte Carlo} (not optical) Glauber simulation~\cite{gluequad,epsnpart}. The Monte Carlo eccentricity $\epsilon_\text{MC}$ rises well above $\epsilon_\text{opt}$ for peripheral and central collisions because of point-like sampling of the nuclear volume. In Fig.~\ref{quadcomp} (right) if $\epsilon_{MC}$ replaces $\epsilon_{opt}$ sharp downturns in the data appear at the centrality extremes (dash-dotted curve) contradicting the possibility of a conjectured ``hydro limit'' for $v_2$ in more-central \auau\ collisions~\cite{volposk}.



\section{Discussion} \label{disc}

We summarize and compare several aspects of jet-related correlations and jet and nonjet quadrupole data systematics and consider alternative interpretations of the nonjet quadrupole in light of inconsistencies in the conventional ``elliptic flow'' hydrodynamic interpretation.

\subsection{2D model fits compared to other $\bf v_2$ methods}

The azimuth quadrupole $A_Q\{2D\}$ or $v_2\{2D\}$ data used in this study are obtained from 2D model fits to angular correlations, but the majority of published $v_2$ measurements are obtained from various NGNM (e.g., $v_2\{2\}$, $v_2\{4\}$, $v_2\{EP\}$). It could be argued that the 2D fit model is somehow arbitrary, that it ``depends on assumptions'' (e.g., choice of model function for the SS 2D peak), and that the inferred quadrupole amplitudes are then uninterpretable and should be ignored~\cite{eplane}. But such arguments imply that NGNM aren't based on model fits and don't ``depend on assumptions.''

Detailed study of the NGNM reveals direct algebraic connections between such methods and 2D angular correlation histograms, and therefore 2D model fits to those histograms~\cite{flowmeth,gluequad,davidhq2}. In fact, most NGNM are actually based on cosine model fits to 2D angular correlations projected onto 1D azimuth while subjected to various conditions on accepted particle pairs, including constraints on $\eta$ difference acceptance and charge combination. 

The NGNM fit model is effectively a single cosine which {\em cannot accurately describe} the 1D projection. The fit residuals are not examined to test the fit validity, and the procedure abandons critical information contained in the unprojected 2D angular correlations. The fitted cosine amplitudes (assumed to represent ``flow'') can then include contributions from multiple correlation components, some identified as ``nonflow.'' Assumptions supporting NGNM methods include identification of any cosine term as representing a flow~\cite{gunther,luzum,multipoles,sextupole} and that ``azimuthal anisotropy'' (any nonuniform structure on azimuth) is dominated by, if not exclusively, flows~\cite{poskvol}.

In contrast, the same 2D fit model defined in Refs.~\cite{davidhq,anomalous} is constrained to describe all 2D data from \pp\ collisions to central \auau\ collisions. The model is based only on observed prominent features of the data, not on physical assumptions. In \pp\ and more-peripheral \auau\ collisions the SS 2D peak is fully resolved and the data {\em require} a SS 2D Gaussian model. The remaining structure (aside from the soft component and BEC) is fully described by two terms of a Fourier series according to inspection of the fit residuals. The two Fourier amplitudes have very different systematic variations on energy and centrality, suggesting minimal parameter covariance. In more-central \auau\ collisions the SS peak persists as a narrow structure on azimuth consistent with a Gaussian.

Systematic uncertainties in the azimuth quadrupole arising from 2D model choices are negligible in the GLS region where the SS 2D peak is fully resolved. In more-central \auau\ collisions it can be shown that the nonjet quadrupole amplitude is insensitive to the SS peak $\eta$ structure {\em as long as a SS Gaussian on azimuth is included in the model}~\cite{tzyam,multipoles,sextupole}. Accurate separation of the three major correlation components is confirmed by the internal consistency of the parameter trends. 
Model comparisons are discussed further in the Appendix.

\subsection{Jet correlation systematics}

Figure~\ref{mike} shows data for the SS 2D peak (left) and AS 1D peak (right) amplitudes scaled by factor ($n_{ch}/N_{bin}$) to determine the ratio of nominally jet-related covariances to number of initial-state \nn\ binary collisions rather than final-state hadrons. The data are plotted vs fractional cross section to emphasize an important point. Within the lower 50\% of the total cross section the SS and AS data agree precisely with binary-collision scaling as expected for dijet production in transparent \auau\ collisions, consistent with the dijet interpretation. The most-probable jets emerge from the lowest-energy partons that can appear as jets in the final state (approximately 3 GeV), as demonstrated in Refs.~\cite{porter2,porter3,ppprd,fragevo,anomalous}. Such low-energy partons  should be most susceptible to a dense, strongly-interacting medium(serving in some sense as ``Brownian probes''~\cite{anomalous}). The data are consistent with no jet modification or medium formation over the more-peripheral half of the total cross section.

Just above the 50\% point (``sharp transition'') the jet-related amplitudes increase substantially relative to the constant GLS trend, the behavior described as ``anomalous centrality variation'' in Ref.~\cite{anomalous}.  But the increase remains consistent with pQCD calculations incorporating modification of fragmentation functions in more-central \auau\ collisions that conserves the parton energy within resolved jets~\cite{fragevo,jetspec}.  The fragment yield increase at lower $p_t$ (e.g.\ 0.5 GeV/c) is precisely anticorrelated with so-called ``jet suppression'' at larger $p_t$ (e.g.\ 10 GeV/c)~\cite{hardspec}.  The jet modification in more-central collisions is not suppression of jet number but rather {\em redistribution} of fragment number along the jet axis from higher \pt\ to lower $p_t$~\cite{fragevo}. 
We conclude that some aspects of parton fragmentation to minimum-bias jets inferred from spectrum analysis and 2D model fits to $p_t$-integral angular correlations remain consistent with a pQCD jet description from \pp\ to central \auau\ collisions.

\subsection{Quadrupole correlation systematics}


In Fig.~\ref{quadcomp} (left panel) we demonstrate precise consistency of 62 and 200 GeV $A_Q\{2D\}$ data scaled by a common $\log(\sqrt{s_{NN}})$ energy dependence shown in Fig.~\ref{energycent} (left panel) similar to that observed for dijets~\cite{ptedep}. Compared to the energy dependence below 13.5 GeV the rate of increase above 13.5 GeV is very large (slope changes by more than a factor 20).  The actual increase in {\em collectivity} below 13.5 GeV is smaller than what the data there suggest due to the kinematic effect of spectator nucleons ( ``sqeezeout'') resulting in negative $v_2$ values at lower energies. Whereas most particles participate in collective motion at lower energies, analysis of $v_2(p_t)$ data to infer ``quadrupole $p_t$ spectra'' at 200 GeV~\cite{davidhq2} suggests that only a small fraction of final-state hadrons participates in the nonjet quadrupole at higher energies~\cite{quadspec,nohydro}. 

In Ref.~\cite{davidhq} it was demonstrated that $A_Q\{2D\}$ data vary approximately as $N_{bin} \epsilon_{opt}^2$. In the right panel $A_Q$ is rescaled as $(n_{ch}/N_{part})(1/N_{bin} \epsilon_{opt}^2)$. The rescaled data are again consistent with a constant value (dashed line) within $\pm 10$\% (hatched band), and the data for two energies are consistent within a few percent modulo the energy scaling factor 1.75, although the absolute $A_Q$ values vary over nearly three decades.


\subsection{Jet-quadrupole comparisons}

By comparing Fig.~\ref{mike} with Fig.~\ref{quadcomp} (left panel) we observe that $A_Q$ attains 60\% of its maximum value within a centrality interval (more-peripheral 50\% of $\sigma/\sigma_0$) that is effectively transparent to jet formation from low-energy (mainly 3 GeV) partons, an interval where multiple (re)scattering of partons or hadrons apparently plays no significant role. Within the transparency interval below the ST we observe that whereas the jet-related covariance scales as $N_{bin}$ (as expected for in-vacuum dijet production in more-peripheral collisions) the nonjet quadrupole covariance scales as $N_{part} \times N_{bin} \times \epsilon_{opt}^2$,
increasing {\em more rapidly than dijet production} modulo the eccentricity factor. 

Figure \ref{quadcomp} (right panel) demonstrates that the nonjet quadrupole continues to follow the same simple algebraic trend within $\pm 10$\% through and above the sharp transition, where the minimum-bias jet trends change dramatically and where substantial modification of jet formation appears~\cite{anomalous}. The quadrupole seems to be completely insensitive to whatever mechanism modifies jet structure.

\subsection{Implications for hydro interpretations}

What are the implications from these observations for hydro interpretations of the azimuth quadrupole? $v_2$ measurements have been conventionally interpreted in a hydro context in terms of ratio $v_2 / \epsilon$ plotted {\em vs} low-density limit (LDL) parameter $(1/S) dn_{ch}/ d\eta$~\cite{volposk,voloshin} ($S$ is the A-A overlap area). For more-peripheral collisions it is expected that $v_2 / \epsilon \propto (1/S) dn_{ch}/ d\eta$ (assumed correlated with the mean number of particle rescatterings during equilibration). If thermal equilibrium is achieved the {\em ideal-hydro limit} $v_2 / \epsilon \rightarrow$ constant (saturation) is expected. 
Previous $v_2$ measurements were believed to confirm that central \auau\ collisions at 200 GeV achieve the ideal-hydro limit (thermalization over some substantial space-time volume)~\cite{uli,voloshin}. 

Conventional $v_2$ analysis is based on assumptions that (a) hydro expansion with particle rescattering is the dominant dynamical process in heavy ion collisions~\cite{ollitrault,uli}, (b) the collision can be described in part as a thermodynamic state~\cite{hydro3}, and (c) $v_2$ is sensitive to an equation of state~\cite{hydro,hydro2}. $v_2$ is defined accordingly~\cite{ollitrault,poskvol}, and nonflow contributions to $v_2$ are estimated using physical-model-dependent procedures~\cite{borg,starv24,2004}. 
The present analysis presents accurate nonjet quadrupole amplitudes derived from physical-model-independent 2D fits to angular correlations that reveal simple trends on centrality and collision energy, including factorization of  the dependence on collision parameters $b$ (impact parameter) and $\sqrt{s_{NN}}$. 

\begin{figure}[h]
  \includegraphics[width=1.65in,height=1.66in]{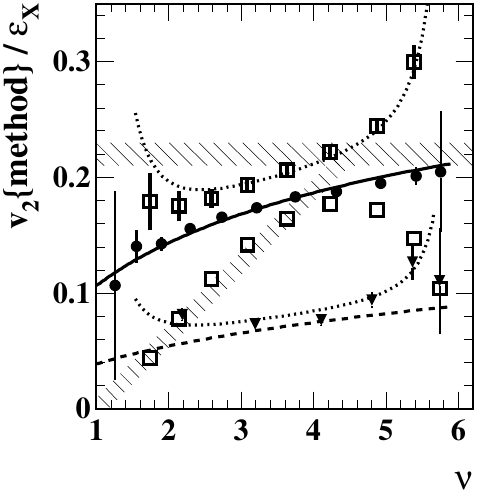}
  \put(-23,30) {\bf (a)}
  \includegraphics[width=1.65in,height=1.66in]{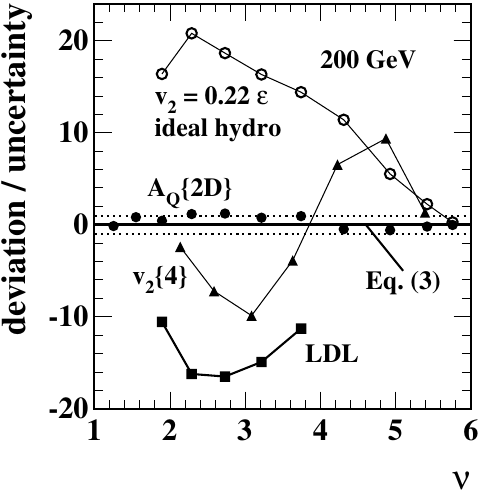}
  \put(-23,30) {\bf (b)}
\caption{\label{fig4}
(a) Ratios $v_2/\epsilon$ for various combinations of $v_2\{method\}$ and $\epsilon_X$ with $X =$ opt or MC. The solid and dashed curves are Eq.~(\ref{loglog}) for 200 and 17 GeV. $v_2\{2\}$ measurements for 200 GeV are open squares~\cite{2004} [published uncertainties (bars) are multiplied by factor 20 for visibility]. $v_2\{EP\}$ measurements for 17 GeV are solid triangles~\cite{na49}. The dotted lines are Eq.~(\ref{loglog}) plus a calculated contribution from the jet-related SS 2D peak.
(b)  Deviations [data - Eq.~(\ref{loglog})] relative to estimated systematic uncertainties for 200 GeV 2D model-fit measurements (solid dots), and for $v_2\{4\}$ (solid triangles), ideal-hydro $v_2/ \epsilon = 0.22$ (open circles) and LDL $v_2/ \epsilon = 0.01\,(dn_{ch}/d\eta \,S)$ (solid squares). $v_2\{method\}$ values have been converted to $A_Q = \rho_0 v_2^2$. 
}
 \end{figure}

Figure~\ref{fig4} (left panel) shows $v_2/\epsilon$ for several $v_2$ methods ($v_2\{2\}$ open squares, $v_2\{EP\}$ solid triangles and $v_2\{2D\}$ solid dots) and two eccentricity $\epsilon$ calculations ($\epsilon_{opt}$ optical and $\epsilon_{MC}$ Monte Carlo). The vertical scale choice excludes a $v_2\{2\}/\epsilon_{opt}$ point at 0.55 for central \auau\ to retain sufficient resolution for the other data. The data points and line types are consistent with Fig.~\ref{v2ep} (left panel). The solid curve is Eq.~(\ref{loglog}) including $\epsilon_{opt}$. The upper dotted curve is the solid curve plus the contribution from the SS 2D (jet) peak calculated in the Appendix. The dashed curve represents Eq.~(\ref{loglog}) for 17 GeV with $\epsilon_{opt}$, and the lower dotted curve is that plus the jet contribution from 200 GeV scaled down by factor $R(\text{17 GeV})$. The bars in the upper open squares represent the published $v_2\{2\}$ systematic uncertainties multiplied by factor 20. The lower open squares are the same $v_2\{2\}$ measurements combined with $\epsilon_{MC}$. The upper hatched band represents an ideal-hydro saturation limit predicted for 200 GeV. The lower (diagonal) hatched band sketches the conjectured LDL trend on $\nu$. Incorporation of $\epsilon_{MC}$ in the $v_2/\epsilon$ ratio partially compensates for the jet contribution to $v_2\{2\}$ and seems to meet LDL expectations for more-peripheral collisions. But the accompanying downturn for more-central collisions contradicts hydro expectations for saturation. Note  that for transparent \aa\ collisions ($\nu < 3$) the LDL trend should be $v_2 / \epsilon = 0$.

Figure~\ref{fig4} (right panel) compares data and theoretical expectations in the form of deviations from Eq.~(\ref{loglog}) divided by data systematic uncertainties (``error'' bars). The $v_2\{2D\}$ data (solid dots) are systematically consistent with Eq.~(\ref{loglog}) as expected. The $v_2\{4\}$ measurement deviations relative to their published uncertainties~\cite{starv24} reveal systematic deviations from Eq.~(\ref{loglog}) of either sign by up to ten uncertainty bars. The ideal-hydro $v_2/\epsilon = 0.22 $ and LDL-scaling $v_2 / \epsilon \approx 0.01\,(dn_{ch}/d\eta \,S)$ also exhibit large deviations (relative to uncertainties in $v_2\{2D\}$ data). 

The more-peripheral NJ quadrupole data do not appear to follow an LDL trend or require subsequent \auau\ collision evolution by particle (parton or hadron) rescattering, and no transition to an ideal-hydro limiting case is observed. The $A_Q\{2D\}$ data do not compel a model description based on bulk-medium hydrodynamics or an equation of state, in fact contradict such descriptions. 

Thus, the measured $A_Q\{2D\}$ data trends, especially the entire  energy-dependence trend at and above Bevalac energies, the insensitivity of the nonjet quadrupole to jet modifications in more-central \auau\ collisions and detailed understanding of jet-related contributions to some $v_2\{\text{method}\}$ measurements (i.e.\ required corrections for jet-related ``nonflow'' bias) strongly suggest that the conventional hydro interpretation of the nonjet quadrupole as elliptic flow is contradicted by most $v_2$ measurements.

\subsection{Hydrodynamic flows vs alternative mechanisms}

Other analysis results argue against hydrodynamic flows in high energy nuclear collisions~\cite{nohydro}. Published $v_2(p_t)$ measurements for identified hadrons reveal a {\em quadrupole $p_t$ spectrum} common to three hadron species and consistent with a boosted source (a form of ``radial flow'') but with a cold spectrum shape very different from that for the majority of final-state hadrons~\cite{quadspec}. The inferred boost distribution is also inconsistent with Hubble expansion of a flowing bulk medium~\cite{davidhq2}. The combined spectrum characteristics suggest that the fraction of hadrons ``carrying'' the nonjet quadrupole is substantially less than 10\%, ruling out a flowing bulk medium as the common source for most hadrons~\cite{quadspec}. The systematics of $v_2\{2D\}(p_t,b)$ data reveal that the source boost does not depend on \auau\ centrality as one might expect for a hydro scenario~\cite{davidhq2}. 

Differential study of single-particle $p_t$ spectra for identified hadrons reveals that spectrum structure conventionally interpreted (with a ``blast-wave'' spectrum model) as representing radial flow~\cite{blast} is actually consistent with parton fragmentation to jets for all \auau\ centralities~\cite{hardspec,fragevo,nohydro}. Mean-$p_t$ systematics from several collision systems at the LHC are consistent with dijet production as the dominant, if not exclusive, source of $\langle p_t \rangle$ variation with \pp\ multiplicity or \aa\ centrality~\cite{alicempt,tommpt}. The nonjet quadrupole increases to a large amplitude within the ``transparent'' centrality interval of \auau\ collisions, with nonzero values down to \nn\ collisions~\cite{davidhq,anomalous}. The measured \nn\ quadrupole systematics extrapolated to LHC energies explain the same-side ``ridge'' in 7 TeV \pp\ collisions as a quadrupole manifestation~\cite{ppcms,lannypp}. Thus, the nonjet quadrupole amplitude takes on large values in collision systems where particle densities are still small, again contradicting a hydro scenario.
If the nonjet quadrupole does not represent ``elliptic flow'' what is the alternative mechanism?  Recent studies suggest that the NJ quadrupole is a QCD phenomenon arising from small-$x$ glue-glue interactions leading to long-wavelength (multipole) QCD radiation~\cite{gluequad}. For example, a QCD calculation of interfering radiation from decays of two BFKL ladders predicts a long-range quadrupole structure in p-p and A-A collisions~\cite{levin}.

Although the centrality trends for jet-related SS peak properties and nonjet quadrupole are very different the two amplitudes, when measured with statistically equivalent quantities $A_{\rm Q}$ and $A_{2D}$, share similar $\log(\sqrt{s_{\rm NN}})$ energy dependences characteristic of QCD scattering processes. Equation~(\ref{loglog}) reveals that the final-state nonjet quadrupole amplitude for (some small fraction of) produced hadrons is simply determined by initial-state parameters $(\sqrt{s_{NN}},b$)  over a large kinematic domain including \nn\ (p-p) collisions. There is no evidence for quadrupole sensitivity to intermediate processes (multiple rescattering, formation of a thermodynamic state, whatever mechanism modifies jet-related
correlations above the ST) within \aa\ collisions. And a prediction $v_2 \approx 0.02$ for pions from 200 GeV \pp\ collisions based on a QCD  color-dipole model~\cite{boris} is consistent with the observed \nn\ limit $v_2\{2D\} \approx 0.02$ in Fig.~\ref{quadfits}~\cite{davidhq,anomalous}.




 \section{Summary} \label{summ}

In summary, 2D angular correlation data for \auau\ collisions at 62 and 200 GeV are employed to re-examine interpretations of the azimuth quadrupole as ``elliptic flow'' conventionally represented by symbol $v_2$, a hydrodynamic response to the eccentricity of the initial-state \aa\ overlap region. Unlike most conventional $v_2$ methods two-dimensional (2D) model fits to angular correlation data distinguish accurately between jet-related features and a nonjet quadrupole represented by symbol $A_Q$.

The nonjet (NJ) quadrupole exhibits simple systematic trends on collision centrality and energy. The trend $A_Q \propto R(\sqrt{s_{NN}})\, N_{bin}(b)\, \epsilon_\text{opt}^2(b)$ with $R(\sqrt{s_{NN}}) \propto \log(\sqrt{s_{NN}} / \text{13.5 GeV})$ accurately describes NJ quadrupole data over a broad range of energies and all \auau\ centralities. All $p_t$-integrated NJ quadrupole data from 17 to 200 GeV are fully described by two \aa\ initial-state parameters. 

In contrast, jet-related features exhibit Glauber linear superposition (GLS) trends (\aa\ transparency) over the more-peripheral 50\% of the \auau\ total cross section, consistent with unmodified dijet production proportional to \nn\ binary collisions. In more-central collisions jet-related amplitudes increase relative to the GLS trend but in a manner still consistent with pQCD when a simple alteration of fragmentation  leading to jets is introduced.  

Within the \aa\ transparency interval minimum-bias (mainly 3 GeV) jet characteristics indicate that parton or hadron rescattering that might lead to hydrodynamic phenomena in a dense medium is negligible.
But within the same centrality interval nonjet quadrupole amplitude $A_Q$ increases to 60\% of its maximum value.

Comparison of conventional $v_2$ measurements with jet-related and nonjet-quadrupole systematics reveals substantial bias in some $v_2$ measurements due to jet-related contributions, the amount depending on the $v_2$ method invoked. Comparison of the energy and centrality trends inferred for the $p_t$-integral nonjet quadrupole with hydro expectations for ``elliptic flow'' reveals substantial disagreement. The ratio $v_2 / \epsilon$ does not follow the number of in-medium rescatterings during equilibration (LDL scaling) for more-peripheral \aa\ collisions and does not transition to a near-constant ideal-hydro trend for more-central collisions and larger collision energies (ideal hydro limit). Quadrupole $p_t$ spectra inferred from identified-hadron $v_2(p_t)$ data are quite different from the spectra for most hadrons. The quadrupole-source boost distribution inferred from such measurements is inconsistent with Hubble expansion of a flowing bulk medium.

We conclude that: (a) NJ quadrupole and jet-related structures can be distinguished accurately by 2D model fits and do arise from two different mechanisms. (b) The NJ quadrupole appearing in more-peripheral (``transparent'') \auau\ collisions cannot arise from a hydro mechanism relying on multiple rescatterings. (c) The same NJ quadrupole is insensitive to any mechanism or environment (dense medium) that might modify jet structure as observed in more-central \aa\ collisions. (d) Both jets and NJ quadrupole exhibit a similar $\log(\sqrt{s_{NN}})$ energy dependence, with threshold near 10 GeV, suggesting a common QCD framework for both phenomena. (e) The NJ quadrupole is not a hydrodynamic flow manifestation.


This material is based upon work supported by the U.S.\ Department of Energy Office of Science, Office of Nuclear Physics under Award Number DE-FG02-97ER41020.


\begin{appendix}

\section{Flows and nonflows} \label{nnonflow}

``Elliptic flow'' ($v_2$) measurements rely on an assortment of analysis methods that encounter the common problem of distinguishing ``flow'' from ``nonflow.'' The present study demonstrates that jet structure can be distinguished accurately from a nonjet remainder that might represent a hydrodynamic flow if flows were relevant to high-energy nuclear collisions~\cite{gluequad,nohydro,anomalous,davidhq,davidhq2}. The distinction is achieved by 2D model fits to angular correlations. Some have argued that such model fits can be dismissed because they ``depend on assumptions'' and the results are therefore arbitrary. 
But (a) no physical assumptions motivated the fit model described by Eqs.~(\ref{eq1}) and (\ref{estructfit}) ~\cite{axialci,anomalous}, and (b) NGNM measurements {\em actually rely on 1D model fits based on physical assumptions}.

The NGNM $v_2$ fit model is a cosine (or cosine plus constant) applied to a 1D projection onto azimuth of {\em all} 2D angular correlations. Different $v_2$ methods are distinguished by the conditions imposed on accepted hadron pairs (charge combination, hadron species, $\eta$ acceptance) in attempts to reduce ``nonflow'' bias to $v_m$ based on {\em physical} assumptions. The fit residuals are not presented (but see comments on ZYAM subtraction below). Based on results from Ref.~\cite{anomalous} the 2D residuals for a 1D single-cosine model must be large.  Some fraction of the jet structure {\em must} be included in NGNM $v_2$ measurements  as a ``nonflow'' bias. The amount of jet-related bias depends on the $v_2$ method. The bias can be predicted accurately  from 2D model fits if $v_2\{method\}$ is sufficiently well defined (e.g.\ the pair $\eta$ acceptance is specified)~\cite{davidhq,davidhq2}.

\subsection{System A vs System B}

We can identify two descriptive systems. System A is based on the observation that within all 2D angular correlation data three prominent features or components labeled (a), (b) and (c) consist of a SS 2D peak, an AS dipole and an azimuth quadrupole not associated with (a). Those features persist for all \aa\ collision systems. We measure their characteristics with 2D model fits. No physical assumptions motivate that description.

pQCD provides the standard description of high energy nuclear collisions, a falsifiable theory that makes real predictions about what should be observed. Included in those predictions is the appearance of (a) and (b) in 2D angular correlations and their systematic properties. The correspondence between predictions and data {\em requires} interpretation of (a) and (b) as pQCD jet-related. The jet interpretation arises from and relies on established QCD as a physical theory. 
What remains is (c), the nonjet quadrupole distinguished from (a) and (b) in all
cases. Its interpretation is questioned. No falsifiable theory
currently predicts all measured properties of the NJ quadrupole. Thus, System A can be represented by  Data = pQCD ``jets'' + ``nonjets'' and is the basis for Ref.~\cite{anomalous} reporting minijet systematics.

System B is based on the primary assumption that ``flows'' (collective motion shared by many particles) must play a major role in high energy nuclear collisions. But ``flows'' are not required to exist at higher energies by QCD. The nucleons participating in collective motion as observed at the Bevalac are no longer relevant. High-energy collisions are dominated by the small-x gluons in the projectiles -- either liberated in place to form ``soft'' hadrons or undergoing ``hard'' scattering to form dijets. 

In System B ``flows'' are associated with one or more cylindrical multipoles. Various NGNM analyses {\em based on physical assumptions} are developed to extract  multipole amplitudes $v_m$ interpreted to represent flows. The methods rely on a common assumption that ``flows'' can be modeled by cosines.  Most NGNM methods (the fit models) do not recognized the existence of a SS 2D (jet) peak in correlation data or the contributions that the SS peak must make to NGNM multipole amplitudes $v_m$. But the SS peak is the dominant source of ``nonflow'' in published $v_m$ measurements. The system of $v_m$ methods and assumptions is complex and changeable. System B is represented by Data = ``flow'' + ``nonflow'' + ``other'' (``other'' represents structure independent of the $v_m$ ``flow'' multipole of interest) and is the basis for analysis in Ref.~\cite{eplane} and other dihadron correlation studies based on {\em  ZYAM subtraction}~\cite{tzyam}. 

\subsection{Mapping from System A to System B}

If a ``flow'' analysis method is sufficiently well-defined we can establish a quantitative relation between systems A and B and demonstrate that the cross terms are large for the usual ``flow'' methods. The ``jet'' component from A is split between ``flow,''  ``nonflow'' and ``other'' in B. The ``nonjet'' component is also split, and some of ``nonjets'' may appear in ``nonflow.'' There is no justification for assuming that ``nonflow'' + ``other'' includes all of ``jets,'' but that common assumption is the basis for ZYAM subtraction~\cite{tzyam,eplane}.

All angular correlation data include a SS 2D peak [feature (a), part of ``jets'']. The SS peak is always narrow on $\phi$ but is elongated on $\eta$ in more-central \auau\ collisions~\cite{anomalous}. 
The Fourier amplitudes for given peak width are represented by factor $F_m(\sigma_{\phi_\Delta})$ for the $m^{th}$ Fourier term (cylindrical multipole)~\cite{tzyam}. Projection of the 2D peak onto 1D azimuth depends on its $\eta$ width relative to the $\eta$ acceptance $\Delta \eta$ and is represented by factor  $G(\sigma_{\eta_\Delta},\Delta \eta)$~\cite{jetspec}. The jet-related quadrupole amplitude derived from SS 2D peak properties is then defined by~\cite{multipoles}
\bea \label{aqss}
2 A_Q\{SS\}(b)  &=&  F_m(\sigma_{\phi_\Delta})G(\sigma_{\eta_\Delta},\Delta \eta) A_{2D}(b). 
\eea
With that expression we can relate nonjet $A_Q\{2D\}$ and jet-related $A_Q\{SS\}$ quadrupole amplitudes in System A to ``flow'' and ``nonflow'' in System B. The expression for more-complex $\eta$-exclusion cuts is derived in Ref.~\cite{multipoles}.

\begin{figure}[h]
  \includegraphics[width=1.65in,height=1.64in]{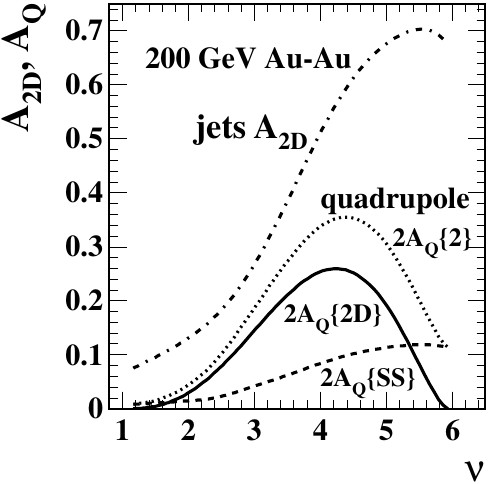}
  \includegraphics[width=1.65in,height=1.66in]{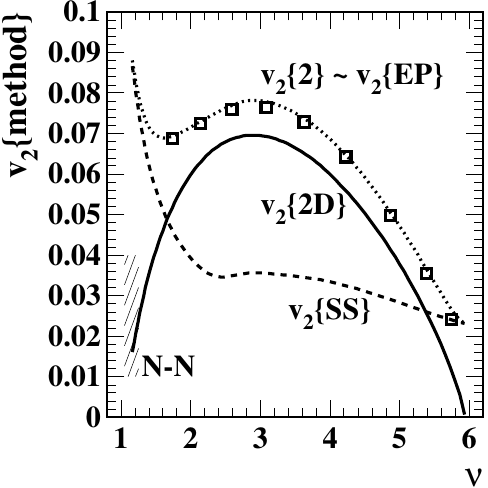}
\caption{\label{nonflow}
(a) SS 2D (jet) peak amplitude $A_{2D}$ (parametrization of data from Ref.~\cite{anomalous}), SS 2D peak quadrupole component $A_Q\{SS\}$ (``nonflow,'' inferred from Ref.~\cite{anomalous} data) and nonjet quadrupole amplitude $A_Q\{2D\}$ from Eq.~(\ref{loglog}), with $A_Q\{2\} = A_Q\{2D\} + A_Q\{SS\}$~\cite{gluequad}. 
(b) Quadrupole amplitudes $A_Q\{X\}$ converted to conventional measures $v_2\{X\}$. The open squares are $v_2\{2\}$ measurements from Ref.~\cite{2004}. The curves correspond to those in the left panel.
 } 
 \end{figure}

Figure~\ref{nonflow} (left panel) shows a parametrization of the centrality dependence of SS 2D peak amplitude $A_{2D}(b)$ and three quadrupole amplitudes  related by~\cite{gluequad}  
\bea
A_Q\{2\} = A_Q\{2D\} + A_Q\{SS\}.
\eea
$A_Q\{2D\}$ (solid curve) is defined by Eq.~(\ref{loglog})~\cite{davidhq} and $A_Q\{SS\}$ (dashed curve) by Eq.~(\ref{aqss}) using SS peak parameters (amplitude and widths) from Ref.~\cite{anomalous}. The sum $A_Q\{2\}$ (dotted curve) is then a {\em prediction} for published $v_2\{2\} \approx v_2\{EP\}$ measurements that are derived from cosine fits to 1D projection onto azimuth of all 2D angular correlation structure (``flow'')~\cite{flowmeth,2004}. The dotted curve in the left panel appears (transformed) in the right panel and in Fig.~\ref{v2ep}.

Figure~\ref{nonflow} (right panel) shows $v_2\{{\rm X}\}$ trends obtained from the corresponding $A_Q\{{\rm X}\}$ curves in the left panel by $A_Q = \rho_0 v_2^2$. Also included are $v_2\{2\}$ measurements (open squares) from Ref.~\cite{2004}. An equivalent comparison for 17 GeV measurements appears in Fig.~\ref{v2ep}. The precise agreement between measurements (points) and prediction (dotted curve) is evident. From the nonjet quadrupole trend and jet-related correlation structure in System A we accurately predict $v_2\{EP\} \approx v_2\{2\}$ published ``flow'' measurements in System B. The prediction does not include small contributions to $v_2\{2\}$ from BEC and electron pairs that are excluded from the SS peak $A_{2D}(b)$ data by the 2D model-fit procedure~\cite{anomalous}.  For statistically well-defined $v_2$ methods (e.g., $v_2\{2\} \approx v_2\{EP\}$) the large ``nonflow'' (jet) bias contribution to inferred ``flow'' $v_2$ can be estimated precisely.

Attempts have been made to parameterize ``nonflow'' contributions to $v_2$ with quantity $g_2 = N_{part}[v_2^2\{2\} - v_2^2\{4\}]$~\cite{2004}. If we approximate $v_2\{4\}$ by $v_2\{2D\}$ we obtain $g_2 = (N_{part}/\rho_0) A_Q\{SS\}\approx \{5/[1+0.1(\nu-1)]\}A_Q\{SS\}$. 
 The factor in curly brackets is derived from a two-component model of particle production~\cite{kn}. That expression agrees quantitatively with 200 GeV $g_2$ data in Fig.~31 of Ref.~\cite{2004} except for more-peripheral collisions where $A_Q\{SS\}$ derived from 2D model fits does not include a relatively large contribution to $v_2\{2\}$ from BEC and electron pairs. The comparison confirms the large ``nonflow'' contribution to $v_2\{2\}$ from the SS jet peak.



\subsection{Consequences of System B}

System B is the basis for dihadron correlation analysis on 1D azimuth including ``ZYAM subtraction'' of a combinatoric background. A ``flow'' background is estimated based on NGNM $v_2$ measurements and subtracted from Data (``raw'' correlations).  It is assumed that for some combination of $v_2$ methods the difference Data $-$ ``flow'' = ``nonflow'' + ``other'' includes all of ``jets.'' The subtraction does reveal the large residuals of the System B model fit. After application of ``trigger-associated'' $p_t$ cuts it is assumed that the surviving ``nonflow'' + ``other''  still retains all of ``jets.'' Since ``flow'' actually includes some fraction of ``jets'' (``nonflow'') the surviving ``jets'' structure in the ZYAM-subtracted and $p_t$-cut ``nonflow'' correlations is attenuated and distorted, leading to incorrect inferences about jet systematics~\cite{tzyam}.  The ``nonflow'' + ``other'' component in System B may include some fraction of ``jets'' from System A, but the fraction depends on arbitrary definitions of ``flow'' and ``nonflow.''

Reference~\cite{eplane} presents a direct comparison between System A and System B. The ZYAM subtraction (System B) shown in Figs.~3 and 7 seems to indicate that with increasing angle relative to the event plane from in-plane to out-of-plane the jet structure is increasingly attenuated and distorted, implying that ``jet quenching'' is directly correlated with the apparent parton path length in a ``dense medium.'' What survives ZYAM subtraction is then further separated into ``jet-like'' and ``ridge-like'' structure, again based on physical assumptions. The paper concludes ``...high $p_t$ triggered jets are biased toward surface emission, and the jet fragmentation is hardly modified by the medium'' (i.e.\ jets in central \auau\ collisions are the same in structure and abundance as in \pp\ collisions).

But Fig.~4 of that paper presents results from 1D model fits (System A), {\em including a SS peak model}, that reveal {undistorted jet structure}. In contrast to severe jet attenuation and distortion with increasing angle inferred from System B the System A model fits reveal a possible {\em increase} of jet correlation amplitude and no distortions, consistent with the minijet analysis in Ref.~\cite{anomalous} in which minijets are modified in more-central collisions, but in a manner consistent with pQCD expectations~\cite{hardspec,fragevo,jetspec}. The System A results in Fig.~4 of Ref.~\cite{eplane} are dismissed because $v_2$ systematics inferred from the 1D model fits contradict ``flow'' $v_2$ inferred from NGNM analysis.

Any application of $v_2$ measurements to subsequent analysis (e.g.\ background subtraction) requires a choice between (a) a comprehensive 2D fit model motivated by actual data structures combined with a few physical-model-independent principles and (b) an incomplete cosine model (NGNM) combined with several  physical-model-dependent {\em a priori} assumptions. Choosing (a) may lead to interesting new physics derived within an intact QCD context representing a falsifiable theory. Choosing (b) (e.g., ZYAM subtraction) can lead to distorted and misleading results.

\end{appendix}


\end{document}